\documentclass[twocolumn]{aastex62}

\usepackage{longtable}
\usepackage{graphicx}
\usepackage{multirow}
\usepackage{color}
\usepackage{amsmath}
\usepackage{booktabs} 
\usepackage{array}
\usepackage{enumerate}
\usepackage[utf8]{inputenc}

\newcommand{\Ha}{\mbox{H$\alpha$}}      
\newcommand{\Hb}{\mbox{H$\beta$}}       
\newcommand{\Hg}{\mbox{H$\gamma$}}      
\newcommand{\HII}{\mbox{\ion{H}{2}}}    
\newcommand{\SII}{\mbox{\ion{S}{2}}}    
\newcommand{\OIII}{\mbox{\ion{O}{3}}}   
\newcommand{\OII}{\mbox{\ion{O}{2}}}    
\newcommand{\NII}{\mbox{\ion{N}{2}}}     

\received{}
\revised{}
\accepted{}
\submitjournal{}

\begin{document}

\title{Physical Properties of {\HII} Regions in M51 from Spectroscopic Observations }

\correspondingauthor{Hu Zou}
\email{zouhu@nao.cas.cn, weipeng@xao.ac.cn}
\author{Peng Wei}
\affil{Xinjiang Astronomical Observatory, Chinese Academy of Sciences, Urumqi 830011, China}
\affiliation{School of Astronomy and Space Science, University of Chinese Academy of Sciences, Beijing 101408, China}
\affiliation{CAS Key Laboratory of Optical Astronomy, National Astronomical Observatories, Chinese Academy of Sciences, Beijing 100101, China}
\affiliation{Center for Astronomical Mega-Science, Chinese Academy of Sciences, 20A Datun Road, Chaoyang District, Beijing, 100012, P. R. China}
\author{Hu Zou}
\affiliation{CAS Key Laboratory of Optical Astronomy, National Astronomical Observatories, Chinese Academy of Sciences, Beijing 100101, China}
\affiliation{Center for Astronomical Mega-Science, Chinese Academy of Sciences, 20A Datun Road, Chaoyang District, Beijing, 100012, P. R. China}
\author{Xu Kong}
\affiliation{Key Laboratory for Research in Galaxies and Cosmology, Department of Astronomy, University of Science and Technology of China, Hefei 230026, China}
\author{Xu Zhou}
\affiliation{CAS Key Laboratory of Optical Astronomy, National Astronomical Observatories, Chinese Academy of Sciences, Beijing 100101, China}
\affiliation{Center for Astronomical Mega-Science, Chinese Academy of Sciences, 20A Datun Road, Chaoyang District, Beijing, 100012, P. R. China}
\author{Ning Hu}
\affiliation{Key Laboratory for Research in Galaxies and Cosmology, Department of Astronomy, University of Science and Technology of China, Hefei 230026, China}
\affiliation{School of Astronomy and Space Science, University of Science and Technology of China, Hefei 230026, China}
\author{Zesen Lin}
\affiliation{Key Laboratory for Research in Galaxies and Cosmology, Department of Astronomy, University of Science and Technology of China, Hefei 230026, China}
\affiliation{School of Astronomy and Space Science, University of Science and Technology of China, Hefei 230026, China}
\author{Yewei Mao}
\affiliation{Center for Astrophysics, GuangZhou University, GuangZhou 510006, P. R. China}
\author{Lin Lin}
\affiliation{Shanghai Astronomical Observatory, Chinese Academy of Sciences, Shanghai 200030, China}
\author{Zhimin Zhou}
\affiliation{CAS Key Laboratory of Optical Astronomy, National Astronomical Observatories, Chinese Academy of Sciences, Beijing 100101, China}
\author{Xiang Liu}
\affiliation{Xinjiang Astronomical Observatory, Chinese Academy of Sciences, Urumqi 830011, China}
\author{Shuguo Ma}
\affiliation{Xinjiang Astronomical Observatory, Chinese Academy of Sciences, Urumqi 830011, China}
\affiliation{School of Astronomy and Space Science, University of Chinese Academy of Sciences, Beijing 101408, China}
\author{Lu Ma}
\affiliation{Xinjiang Astronomical Observatory, Chinese Academy of Sciences, Urumqi 830011, China}
\author{Tuhong Zhong}
\affiliation{Xinjiang Astronomical Observatory, Chinese Academy of Sciences, Urumqi 830011, China}
\author{Fei Dang}
\affiliation{Xinjiang Astronomical Observatory, Chinese Academy of Sciences, Urumqi 830011, China}
\author{Jiantao Sun}
\affiliation{Xinjiang Astronomical Observatory, Chinese Academy of Sciences, Urumqi 830011, China}
\author{Xinkui Lin}
\affiliation{Xinjiang Astronomical Observatory, Chinese Academy of Sciences, Urumqi 830011, China}



\begin{abstract}
M51 and NGC 5195 is an interacting system that can be explored in great details with ground-based telescopes. The {\HII} regions in M51 were observed using the 2.16 m telescope of the National Astronomical Observatories of the Chinese Academy of Sciences and the 6.5 m Multiple Mirror Telescope with spatial resolution of less than $\sim100$ pc. We obtain a total of 113 spectra across the galaxy and combine the literature data of Croxall et al. to derive a series of physical properties, including the gas-phase extinction, stellar population age, star formation rate (SFR) surface density, and oxygen abundance. The spatial distributions and radial profiles of these properties are investigated in order to study the characteristics of M51 and the clues to the formation and evolution of this galaxy. M51 presents a mild radial extinction gradient. The lower gas-phase extinction in the north spiral arms compared to the south arms are possibly caused by the past encounters with the companion galaxy of NGC 5195. A number of {\HII} regions have the stellar age between 50 and 500 Myr, consistent with the recent interaction history by simulations in the literatures. The SFR surface density presents a mild radial gradient, which is ubiquitous in spiral galaxies. There is a negative metallicity gradient of $-0.08$ dex $R_{e}^{-1}$ in the disk region, which is also commonly found in many spiral galaxies. It is supported by the ``inside-out"  scenario of galaxy formation. We find a positive abundance gradient of 0.26 dex $R_{e}^{-1}$ in the inner region. There are possible reasons causing the positive gradient, including the freezing of the chemical enrichment due to the star-forming quenching in the bulge and the gas infall and dilution due to the pseudobulge growth and/or galactic interaction.
\end{abstract}



\section{Introduction}\label{sec:intro}

Understanding galaxy formation and evolution is one of the ultimate challenges in extragalactic astronomy. Detailed investigations of the spatial distributions of stars, dust, and gas in galaxies can provide important information of galaxy formation and evolution, such as star formation, chemical enrichment, and mass assembly. Nearby galaxies provide one of the best laboratories for understanding the current star formation process and can be used to constrain the chemical evolution theories of galaxies \citep{osterbrock06,dob10}. 

Due to strong star formation and ionized hydrogen formed by high-energy radiation from young massive stars, {\HII} regions are prefect probes for studying the star formation processes, evolution of young massive stars, and surrounding interstellar medium. Combining the spectroscopic and multi-wavelength photometric data, we   can obtain a series of physical properties from the measurements of the nebular emission lines and underlying stellar continua, such as star formation rate \citep[SFR;][]{lopez10,Zhou14,gonzalez16}, mass and luminosity \citep{rosales12,garca19},  effective yield and rotation velocity \citep{pilyugin04,zou2011a,hu18}, stellar-to-gas fraction \citep{zahid14}, spatial distribution of gas-phase or stellar metallicity \citep{bre04, zou2011b, zou2011c, lin13, sanchez14, pilyugin14, croxall15, ho15, lin17, hu18}, and stellar population parameters \citep{zou2011b, sanchezB14, Zhou14, hu18}. 
 
In the past decades, the integral field spectrograph (IFS) plays an important role in understanding the nature of galactic structure, formation, and evolution. A number of IFS surveys have being carried out, e.g., SAURON\citep{bacon01}, CALIFA\citep{sanchez12}, SAMI\citep{bryant15}, AMUSING\citep{galbany16}, and MaNGA\citep{bundy15}. Spatially-resolved spectroscopic data of galaxies in the local universe make it possible to statistically study the physical properties and their correlations in sub-galactic scale, such as the radial metallicity gradients \citep{sanchez14, pilyugin14, ho15, croxall16, sanchezM19}, radial age gradients \citep{sanchezB14}, the relation between the abundance gradient and morphology, mass, or bar \citep{sanchezM18, zinchenko19}, SFR as a function of Hubble type and of galaxy mass \citep{gonzalez16,cano19}, and global and local mass-metallicity relation \citep{tremonti04,rosales12,barrera16}.

Most IFS surveys focus on the galaxies with relatively small apparent sizes, generating the spatial resolutions of order of kpc or sub-kpc. Considering that nearby large galaxies are rather close to us and their sizes are too big to be covered by IFS observations due to relatively small field of view, we have undertaken a project of spectroscopic observations of {\HII} regions in 20 nearby face-on spiral galaxies \citep{kong14}, using the long-slit spectrograph of the 2.16 m telescope \citep{fan16} mounted at XingLong station of National Astronomical Observatories of China (NAOC) and multifiber spectrograph of the Multiple Mirror Telescope \citep[MMT;][]{fabricant05}. With these samples, we can obtain high spatial-resolution ($<$ 160 pc) spectroscopic data of galaxies to analyze the distributions of the galaxy observables in great details, including dust extinction, metal abundance, star formation, and stellar population. So far, we have obtained the largest spectral sample of {\HII} regions for M33 and analyzed the spatial distribution of electron temperature and oxygen abundance \citep{lin17}. \citet{hu18} used 188 spectrum of {\HII} regions of M101 to investigate the two-dimensional distributions of stellar population and kinematic properties of this galaxy. 

As one of the 20 nearby galaxy samples, M51 has been observed by using both the long-slit of the 2.16 m telescope and multifiber spectorgraph of the MMT. M51 (NGC 5194, also known as the Whirlpool nebula, $\alpha$ =$13^{h}29^{m}52.^{s}711$, $\delta$ = +$47{\arcdeg}11{\arcmin}42.{\arcsec}62$) is a grand-design face-on spiral galaxy with the Hubble type of Sbc. Since M51 and its peculiar companion galaxy NGC 5195 are a close interacting galaxy pair, substantial {\HII} regions have been formed in its spiral arms. M51 presents the metal-rich feature and a shallow radial abundance gradient \citep{bre04,croxall15}. It was found that M51 and NGC 5195 underwent an interaction, which induced a burst of star formation about 340--500 Myr ago \citep{men12}. Therefore, M51 is an excellent object for studying the current status of an interacting system. In this paper, we present the spectroscopic observations of the {\HII} regions and derive a series of physical properties, including the extinction, SFR surface density, stellar age, and gas-phase metallicity, and study their spatial distributions. Through the properties and their spatial distributions, we try to obtain the characteristics of this galaxy and investigate evolutionary clues and possible influence of the galactic interaction. Table \ref{tab:pars} lists some basic parameters for M51 and NGC 5195, which are adopted for analyzing the radial profiles in this paper. 
\begin{deluxetable}{cccc}[!htb]
\tablecaption{Basic parameters for M51 and NGC 5195. \label{tab:pars}}
\tablecolumns{4}
\tablehead{ & \colhead{Parameters} & \colhead{Parameter value} & \colhead{Reference\tablenotemark{a}}}
\startdata 
         & R.A. (J2000)      & $13^{h}29^{m}52.^{s}711$                & (1) \\
         & Decl. (J2000)     & +$47{\arcdeg}11{\arcmin}42.{\arcsec}62$ & (1) \\
         & Distance          & 7.9 Mpc                                 & (1) \\
  M51   & Inclination       & 22\arcdeg                               & (2) \\
         & P.A.              & 172\arcdeg                              & (3) \\
         & $R_{25}$          & 336.{\arcsec}6                          & (4) \\
         & $R_{e}$           & 125.{\arcsec}1                          & (4) \\
\tableline
         & R.A. (J2000)      & $13^{h}29^{m}59.^{s}590$                & (1) \\
         & Decl. (J2000)     & +$47{\arcdeg}15{\arcmin}58.{\arcsec}10$ & (1) \\
NGC 5195 & Distance          & 7.9 Mpc                                 & (1) \\
         & inclination       & 43\arcdeg                               & (5) \\
         & P.A.              & 91\arcdeg                               & (5) \\
\enddata
\tablenotetext{a}{References: (1) NASA/IPAC Extragalactic Database (NED); (2) \citet{colombo14}; (3) \citet{walter08}; (4) Third Reference Catalog of Bright Galaxies (RC3) \citep{de95}; (5) \citet{spillar92}.}
\end{deluxetable}

The paper is structured as follows. Section \ref{sec:data} introduces the spectroscopic observations and corresponding data reduction. Section \ref{sec:measurement} describes the detailed measurements of different physical properties. Sections \ref{sec:result} shows the spatial distributions of these physical properties and presents related comparison and analysis. The summary is given in Section \ref{sec:summary}.

\section{Observations and Data Reduction} \label{sec:data}
\subsection{Facilities and Observations} \label{subsec:obs}
The {\HII} regions were selected from the continuum-subtracted {\Ha} image from NASA/IPAC Extragalactic Database. The sources are detected in this {\Ha} image  by SExtractor \citep{bertin96}. The spectroscopic targets of {\HII} regions are selected as those sources with at least 25 pixels (about 75 pc in radius) whose {\Ha} flux is larger than a critical value. Possible contaminations by foreground bright stars selected from Two Micron All Sky Survey Point Source Catalog \citep{skrutskie06} were eliminated. The optical spectra of M51 were taken by the Optomechanics Research Inc.(OMR) long-slit spectrograph of the 2.16 m telescope \citep{fan16} at the Xinglong Station of NAOC and the Hectospec multi-fiber positioner and spectrograph of the 6.5 m MMT telescope \citep{fabricant05}. The OMR spectrograph is deployed at the Cassegrain focus. It provides a dispersion of 4.8 {\AA}  pixel$^{-1}$ and a resolution of  about 10 \AA. The wavelength coverage is about 3600–8000 \AA. The Hectospec is a moderate-resolution, multi-object optical spectrograph at the Cassegrain focus of the 6.5 m MMT telescope. The Hectospec fiber positioners allow the users to reconfigure 300 fibers to targets over the 1{\arcdeg} focal surface in 300 seconds. Each fiber has a diameter of 1.\arcsec5, corresponding to about 57 pc at the distance of M51. The Hectospec 270 gpm grating with blaze wavelength at 5200 {\AA} was used. The spectral coverage ranges from 3650 to 9200 {\AA}. The dispersion is 1.21 $\mbox{\AA}\ \mathrm{pixel}^{-1}$ and the resolution is $\sim$ 5 {\AA}.

\begin{figure*}[!htb]
\center
\resizebox{\width}{!}{\rotatebox{-90}
{\includegraphics[width=0.55\textwidth]{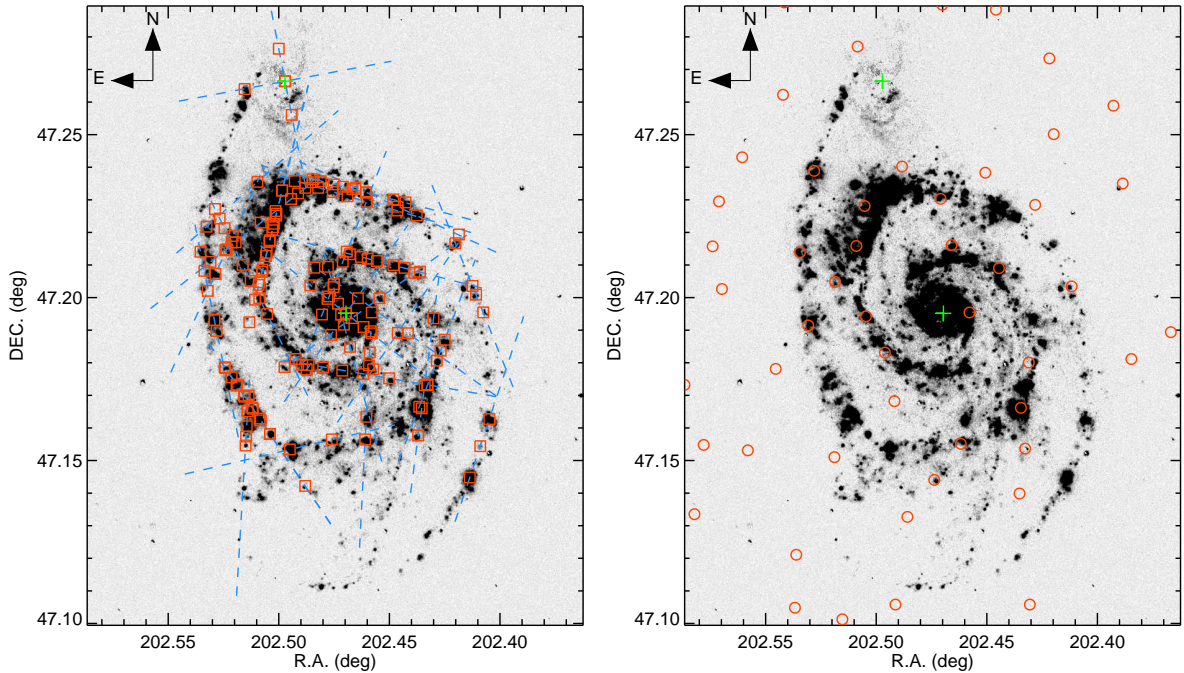}}}
\caption{Left: slit positions (blue dashed line) placed on M51 observed by the NAOC 2.16 m telescope . The background gray-scale map is the continuum-subtracted {\Ha} narrowband image taken from NED. Green pluses imply the centers of M51 and NGC 5195. North is up and east is left. The red squares show the positions of the extracted spectra. Right: fiber positions (open circles) of the Hectospec multi-fiber positioner on the 6.5 m MMT telescope.  }
\label{fig:obs}
\end{figure*}

The M51 observations with the OMR spectrograph of the NAOC 2.16 m telescope started in 2008 March 9 and ended in 2014 March 23. A total of 19 nights were used. Bias and dome flat frames were taken at the beginning and end of every night. The He--Ar arc lamp was used for wavelength calibrations.  Normally, we took two exposures for each object and each was about 1800 s.  The left panel of Figure \ref{fig:obs} presents the slit positions on the sky. The length of the spectrographic slit was about 4{\arcmin} and the width was about 2.\arcsec5, corresponding to about 96 pc at the distance of M51. A total of 30 slit positions were selected. Some of them were placed along the major-axis and minor-axis directions of M51 and NGC 5195, which was aimed to obtain high signal-to-noise ratio (S/N) continua of the bulges. For other slits, we manually placed them to cover as many {\HII} regions as possible. The sky spectra with exposure time of 1200 s were taken between two exposures of each slit position. Proper flux standard stars were selected from the catalog of International Reference Stars \citep[IRS;][]{corbin91}. Their slit spectra were used for the flux calibration. 

We were awarded half a night on 2012 February 10 by the Telescope Access Program (TAP\footnote{http://info.bao.ac.cn/tap/}) to use the multifiber spectrograph of the MMT telescope . The {\HII} regions of M101 were preferentially targeted due to its larger size. M51 was considered as a secondary target to supplement the long-slit observations. It was unfortunate that the weather got worse after the M101 observation. Only a few spectra with good quality were obtained. The exposure time was 3600 s. The observations were taken at airmass of about 1.06 and the seeing was about 0.\arcsec8. The right panel of Figure \ref{fig:obs} shows the fiber positions.

\subsection{Spectral Data Reduction} \label{subsec:reduction}
The raw data taken by the NAOC 2.16 m telescope observations are processed using the IRAF software\footnote{\url{http://iraf.noao.edu/iraf/web/iraf-homepage.html}}. We perform the bias subtraction, flat-field correction, cosmic-ray removal, and the sky-background subtraction. The spectrum of each {\HII} region is extracted based on the dispersion trace of the flux standard star and then wavelength-calibrated by using the extracted spectrum of the He--Ar lamp at the same CCD position.  Flux calibration is performed based on the observation of the flux standard star and the mean atmospheric extinction coefficients at the Xinglong Station.  

The MMT data are reduced with the publicly available HSRED v2.0 software\footnote{http://mmto.org/rcool/hsred/index.html}. After bias subtraction and flat-field correction, the fiber spectra are extracted and wavelength calibrated. The background sky emission is estimated by taking the average spectra with “blank sky” fibers in the same exposure and is subtracted from the individual spectrum. There is no observation for the flux standard star. The spectral energy distributions (SEDs) from the 15 intermediate-band photometric images of the Beijing--Arizona--Taipei--Connecticut (BATC) Color Survey of the Sky are utilized to calibrate the spectra \citep{lin17}. 

There is a total of 250 extracted spectra, 113 of which are visually checked to have either good stellar continua or emission lines. Among the 113 spectra, 99 spectra are from the slits and the others are from the fibers. The furthest location of the spectra was $\sim$ 10.5 kpc in terms of galactocentric distance of M51. All these spectra are resampled (keeping energy conservation) with a wavelength step of 1 {\AA} in the range of 3700–7500 {\AA} using IRAF’s $dispcor$ procedure. The 1 {\AA} step is selected according to the suggestion of STARLIGHT\footnote{http://www.starlight.ufsc.br} spectral synthesis code \citep{cid05}, as we want to get the stellar population age. These spectra are corrected for the Galactic extinction using the extinction law of \citet{car89} and the reddening value of $E(B-V)$ = 0.031 from the Galactic dust map of \citet{schlegel98}.

\section{Spectral Measurements} \label{sec:measurement}

\subsection{Full spectrum fitting and stellar age measurement} \label{subsec:specfit}

\begin{figure}[!htb]
\center
\resizebox{\width}{!}{\rotatebox{-90}
{\includegraphics[width=0.49\textwidth]{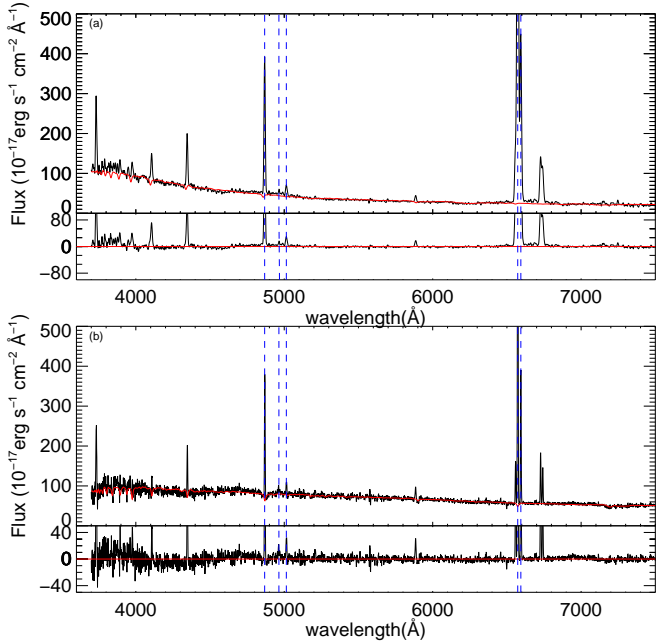}}}
\caption{Upper: a high S/N observed spectrum of an {\HII} region (black in the upper panel) observed by the NAOC 2.16 telescope. The best-fit model spectrum is overplotted in red. The residual spectrum between the observed and model spectra is plotted in the lower panel . The horizontal red line denotes the zero residual. The vertical dashed lines mark the emission lines used in this work. From left to right, they are {\Hb}, [{\OIII}]$\lambda$4959, [{\OIII}]$\lambda$5007, {\Ha}, and [{\NII}]$\lambda$6583, respectively. Bottom: the same as an arbitrarily selected {\HII} region observed by the MMT telescope.}
\label{fig:spectra}
\end{figure}

The underlying stellar continuum of each spectrum is modeled with the STARLIGHT spectral synthesis code. A grid of 150 simple stellar population (SSP) templates from \citet{bru03}\footnote{http://www.bruzual.org} with the \citet{chab03} initial mass function (IMF) are used. These SSP templates cover 25 different ages (1 Myr–18 Gyr) and six metallicities (0.005 -- 2.5 Z$_{\sun}$). The extinction law of \citet{car89} is adopted to add dust extinction to the models. The observed spectrum is considered as a linear combination of the SSP templates. During the fitting of STARLIGHT, the wavelength regions covered by nebular emission lines and atmospheric absorption lines are masked. Figure \ref{fig:spectra} gives an example of the spectral fitting for two {\HII} regions observed by the NAOC 2.16 telescope and MMT. The best-fit model spectra are also overlapped in this figure. 

For each spectrum, we derive a stellar population age, which is a light-weighted average of the SSP compositions. In order to get a relatively reliable age estimate, the continuum S/N at $\sim 5500$ {\AA} is required to be higher than 10. A total of 86 spectra have the age measurements. The uncertainty in logarithmic stellar age estimated by \citet{cid05} is about 0.08 dex for S/N $>$ 10.

\subsection{Emission-line measurements} \label{subsec:lineflux}
After the subtraction of the stellar continuum as modeled in Section \ref{subsec:specfit}, the emission-line fluxes of {\Hb}, [{\OIII}]$\lambda$5007, {\Ha}, and [{\NII}]$\lambda$6583, which are used in this paper, are measured through the Gaussian function fitting. Only a few spectra (e.g. the ones at the galactic cores as shown in Figure \ref{fig:spectra}) have the detections of other emission lines, such as [{\OII}] and [{\SII}], so their flux measurements are not provided. Due to relatively low spectral resolution and low S/N of the spectra, the flux ratio between [{\NII}]$\lambda$6583 and [{\NII}]$\lambda$6548 is tied to theoretical values of 0.348, respectively. The profiles of {\Ha}, [{\NII}]$\lambda$6548, and [{\NII}]$\lambda$6583 are fitted simultaneously. Following \citet{hu18}, the flux uncertainty of each emission line is estimated with the standard deviation within a wavelength region of 200 {\AA} in width at the central wavelength of the emission line and errors induced by the Gaussian fitting. 

\subsection{Extinction estimation} \label{subsec:reddencor}
The gas-phase extinction can be estimated with the Balmer line ratios, such as {\Ha}/{\Hb}, {\Ha}/{\Hg}, and {\Hb}/{\Hg}. The extinction estimation based on the Balmer lines is regarded to be independent on the physical conditions of the gas, such as the volume density and temperature \citep{domnguez13}. The {\Ha} and {\Hb} lines are detected in most of our spectra, so we use {\Ha}/{\Hb} to estimate the dust extinction. 

Based on the assumptions of the reddening law of \citet{car89}, $R_{V}=3.1$, and the intrinsic {\Ha}/{\Hb} of 2.86 under the Case B recombination of $T_{e}$ = 10,000 K and $n_{e}$ =100 cm$^{-3}$\citep{osterbrock06}, the reddening value of $E(B–V)$ can be calculated as
\begin{equation}
E(B-V)=\frac{2.5}{k(\mathrm{H\beta)}-k(\mathrm{H\alpha)}}\log{\frac{(\mathrm{H\alpha}/\mathrm{H\beta})_{\mathrm{obs}}}{(\mathrm{H\alpha}/\mathrm{H\beta})_{\mathrm{int}}}},
\end{equation}
where $k$({\Ha}) and $k$({\Hb}) represent the values of the reddening law at the wavelengths of the {\Ha} and {\Hb} lines, respectively. ({\Ha}/{\Hb})$_\mathrm{obs}$ and ({\Ha}/{\Hb})$_\mathrm{int}$ are the observed and intrinsic flux ratios, respectively. As described in \citet{cat15}, the extinction curves and dust-to-stars geometries would have little effect on the attenuation calculation. We also check the effect of adding the dust curve of stellar content in the SSP fitting on the gas-phase extinction and find that it makes the gas-phase extinction difference within about 0.03 mag, which is quite smaller than the average measurement error of about 0.1 mag. In this paper, if the observed {\Ha}/{\Hb} $\le$ 2.86, $E(B–V)$ is set to zero.  All the fluxes of the emission lines used in this paper are corrected for this gas-phase extinction. 

\subsection{$\Sigma_{\mathrm{SFR}}$} \label{subsec:SFRmethod}
The SFR surface density is calculated as $\Sigma_{\mathrm{SFR}} = \mathrm{SFR}/\mathrm{area}$. The area for a long-slit spectrum is calculated as the one of a rectangle, whose length is the aperture size used for extracting the spectrum and width is the slit width. The area for a fiber spectrum is calculated as the one of an circular, whose diameter is the fiber size. The SFR in $\mathrm{M_{\sun}\ yr^{-1}}$ is estimated with the extinction-corrected {\Ha} luminosity ($L({\Ha})$) by using the relation from \citet{kennicutt98}: 
\begin{equation}
\mathrm{SFR} = 7.9\times10^{-42}L({\Ha})(\mathrm{erg}\ \mathrm{s^{-1}}).
\end{equation}
This SFR calculator is derived based on the Salpeter IMF with stellar masses in the range of 0.1--100 $\mathrm{M_{\sun}}$. Note that the {\Ha} luminosity is calibrated by using the aperture flux of 3{\arcsec} in diameter in the Sloan Digital Sky Surveys $r$-band image.

\subsection{Metallicity Determination} \label{subsec:O/Hmethod}
The so-called $direct$ $T_{e}$ method is the most reliable way to measure gas-phase oxygen abundance, which is based on the ratio of auroral line intensities, such as [{\OIII}]$\lambda$4363/$\lambda$5007 \citep{osterbrock06}. However, it is difficulty to detect these auroral lines in our spectra, which are $\sim$ 100–1000 times fainter than {\Hb}. There are multiple strong-line methods to estimate the gas-phase oxygen abundance of galaxies, including R23 \citep{kobu99,pilyugin05}, N2 \citep{pp04,marino13}, O3N2 \citep{pp04,marino13}, N2O2 \citep{kewley02,bre07}, etc. Since only a few spectra have the detection of [{\OII}]$\lambda$3727, those metallicity estimators related to this line is not applicable. Although [{\NII}]$\lambda$6583 and {\Ha} lines are detected for most of our spectra,  the metallicity based on the N2 index ($\mathrm{N2}$ $\tbond$ $\log$([{\NII}]$\lambda$6583/{\Ha})) reaches saturation in the solar and super-solar metallicity regime with the N2 calibration \citep{pp04}. Thus, we use the O3N2 index to estimate the gas-phase oxygen abundance, where 
\begin{equation}\label{eq:o3n2}
\mathrm{O3N2}=\log{(\frac{[\OIII]\lambda5007/\Hb}{[\NII]\lambda6583/\Ha})}.
\end{equation}

We adopt the O3N2 metallicity calibration of \citet{marino13}, which is expressed as
\begin{equation}
12+\log{(\mathrm{O/H})}=8.533-0.214\times \mathrm{O3N2}.
\end{equation}
This calibration is valid in the range of  -1.1 $<$ O3N2 $<$ 1.7 and the calibration uncertainty is about 0.18 dex \citep{marino13}. The random error of the metallicity, which is provided in the rest of this paper, comes from the emission-line measurements. 

\section{Results} \label{sec:result}
Table \ref{tab:flux} in the Appendix \ref{appe:tab2} shows all the emission line measurements and corresponding spectral properties. A total of 113 spectra with {\Ha} and {\Hb} flux S/N greater than 5 are presented. All the emission-line fluxes are corrected for the gas-phase extinction. The {\Hb} flux is the absolute line strength, while other line fluxes are relative ones, which are  normalized to the {\Hb} flux. The metallicity measurement requires that [\OIII]$\lambda$5007 and [\NII]$\lambda$6583 have S/Ns higher than 5 (a total of 67 regions). The age measurement requires that the continuum S/N at 5500 {\AA} higher than 10 (a total of 86 regions).

\begin{figure}[!htb]
\center
\resizebox{\width}{!}{\rotatebox{-90}
{\includegraphics[width=0.45\textwidth]{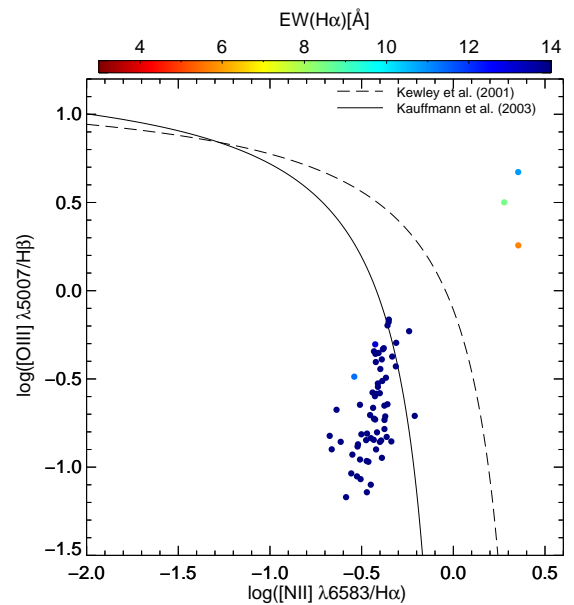}}}
\caption{Our spectral samples on the BPT diagram presenting the [{\OIII}]$\lambda$5007/{\Hb} line ratio against [{\NII}]$\lambda$6583/{\Ha}. The curves with solid and dashed lines are from \citet{kauffmann03} and \citet{kewley01}, respectively. The color of each point indicate the equivalent width of {\Ha} of each spectra.} 
\label{fig:bpt}
\end{figure}

In order to discriminate dominant ionizing sources, we apply the Baldwin–Phillips–Terlevich (BPT) diagram and the equivalent width of {\Ha} (EW({\Ha})) to examine the excitation properties of the spectra. As shown in Figure \ref{fig:bpt}, the BPT diagram presents the [{\OIII}]$\lambda$5007/{\Hb} line ratio against [{\NII}]$\lambda$6583/{\Ha} \citep{baldwin81}. Two lines from \citet{kewley01} and \citet{kauffmann03} are adopted to separate pure star-forming, active galactic nuclei (AGNs), and composite regions. As described in  \citet{lacerda18}, the EW({\Ha}) can be also used to diagnose different ionization sources in a single galaxy: regions where EW({\Ha}) $<$ 3 {\AA} are ionized by hot low-mass evolved stars, regions where EW({\Ha}) $>$ 14 {\AA} are related to the ionization due to young OB stars in {\HII} regions and trace the star formation complexes, and regions where 3 {\AA} $<$ EW({\Ha}) $<$ 14 {\AA} are contributed by more than one process. The EW({\Ha}) of our spectra is shown as the colors of the data points in Figure \ref{fig:bpt}.

In the BPT diagram, the data points above the \citet{kewley01} curve are regarded as the regions possibly affected by AGN. Three points are located in the AGN region and the EW({\Ha}) is smaller than 14 {\AA}. These points are closed to the nucleus of M51, which has the AGN activity as reported by \citet{goad79} and \citet{moustakas10}. These exceptions, which is specially noted in Table \ref{tab:flux}, are excluded in the following analyses. There are additional 2 points with EW({\Ha}) $< 14$ {\AA} . Most of our samples are star-forming regions close to or below the \citet{kauffmann03} curve and their EW({\Ha}) is larger than 14 {\AA}.  All other spectra that have not enough S/Ns of emission lines and thus are not on the BPT diagram are located far away from the M51 nucleus, and their EW({\Ha}) is larger than 14 {\AA}. It is believed that our spectra mainly come from star-forming regions.

\subsection{Extinction distribution} \label{subsec:dust}
\begin{figure}[!htbp]
\center
\resizebox{\width}{!}{\rotatebox{-90}
{\includegraphics[width=0.45\textwidth]{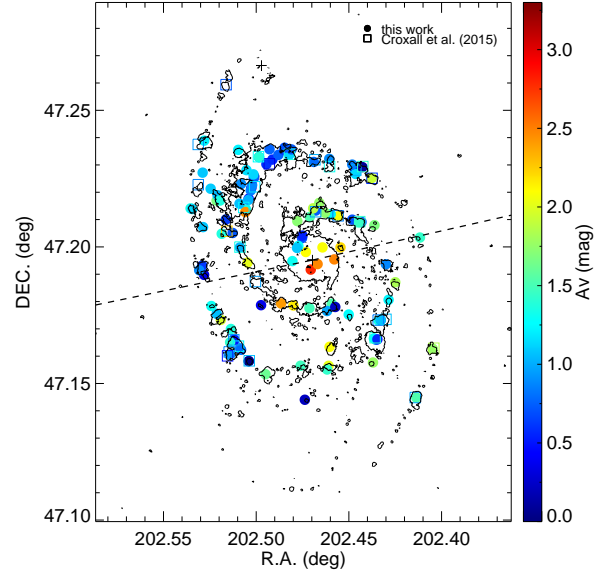}}}
\caption{Two-dimensional distribution of the gas-phase extinction $A_{V}$ across M51. The contours display the isophotal shapes of the {\Ha} emission. The centers of M51 and NGC 5195 are marked with pluses. The filled-circles are the extinctions obtained in this work and the open-squares are the extinctions from \citet{croxall15}. The dashed line shows the direction of the minor axis, which is used as the boundary between the northern and southern parts of M51.}
\label{fig:av}
\end{figure}

Based on the Balmer decrement and assumed reddening law of \citet{car89}, we derive the extinction in $A_V$ and map the extinction distribution in Figure \ref{fig:av}. In this figure, we also present the extinctions derived by \citet{croxall15}, who obtained the spectra of 59 {\HII} regions in M51 from the CHemical Abundances Of Spirals (CHAOS) project \citep{berg15}. In the paper of \citet{croxall15}, the extinction was given at {\Hb}. The {\Hb} extinction c({\Hb}) is converted to $A_{V}$ following c({\Hb}) = 1.43$E(B-V)$ and $A_{V}=3.1 E(B-V)$ \citep{berg15}. The overall average gas-phase extinction is about 1.18 mag. The core region of M51 presents a larger gas-phase extinction than the outside. 

\begin{figure*}[!tb]
\center
\resizebox{\width}{!}{\rotatebox{-90}
{\includegraphics[width=0.35\textwidth]{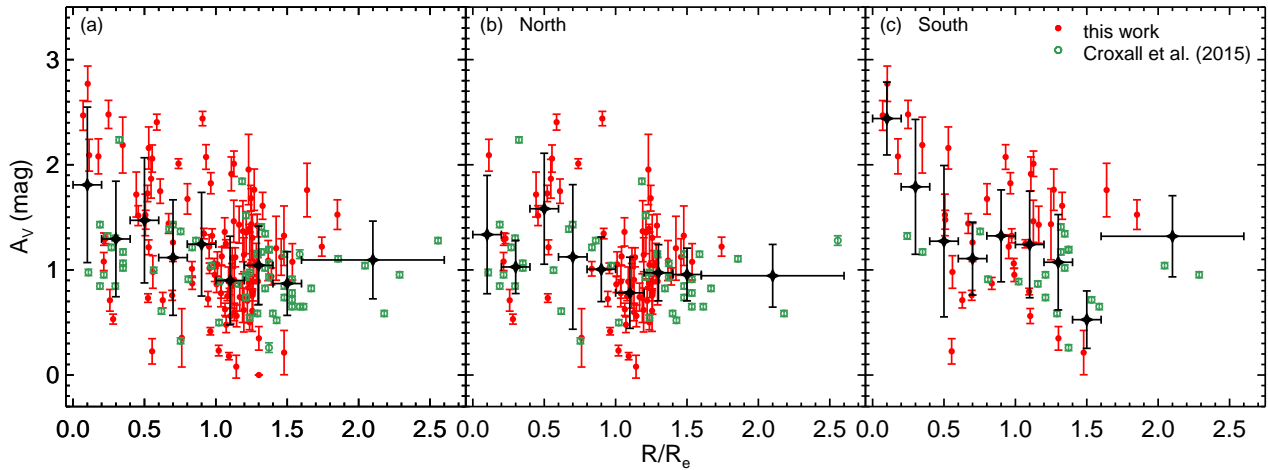}}}
\caption{(a) Deprojected radial extinction distribution of all our {\HII}-region samples in M51. Here $R_e$ is the effective radius of M51, which is shown in Table \ref{tab:pars}. The filled-circles and open-circles are the data from this work and \citet{croxall15}, respectively. The black star with vertical error bar presents the average extinction and its standard deviation in each radial bin, while the horizontal bar shows the range of each bin. (b) Deprojected radial extinction distributions of the north part. (c) Deprojected radial extinction distributions of the south part. }
\label{fig:avgradient}
\end{figure*}

The radial extinction distribution is shown in Figure \ref {fig:avgradient}a. We can see that there is a mild extinction gradient, although the radial extinction presents a relatively large dispersion.  From the extinction map in Figure \ref{fig:av}, we can also see that the extinction in the northern spiral arms is generally smaller than the southern ones. Therefore, we try to divide the spectral samples into north and south regions and present their radial extinction distributions in Figure \ref{fig:avgradient}b and \ref{fig:avgradient}c, respectively. The line of the minor axis of M51 is considered as the boundary, which is plotted in dashed line in Figure \ref{fig:av}. From these radial distributions, we find that the dispersions get reduced compared to the whole distribution and the gas-phase extinction of the north region, which is close to NGC 5195, is generally smaller than that of the south one. The M51--NGC 5195 system has undergone more than once close encounters and the last one possibly occurred 300-500 Myr ago, which was inferred from kinematic and hydrodynamic simulations \citep{salo00, dob10}. The close encounter might disperse the gas distribution in the north parts more seriously and lead to the low gas-phase extinction, although more observational evidence is needed.

\subsection{Stellar age and $\Sigma_{\mathrm{SFR}}$ distributions} \label{subsec:age}
\begin{figure*}[!htb]
\center
\resizebox{\width}{!};{\rotatebox{-90}
{\includegraphics[width=0.34\textwidth]{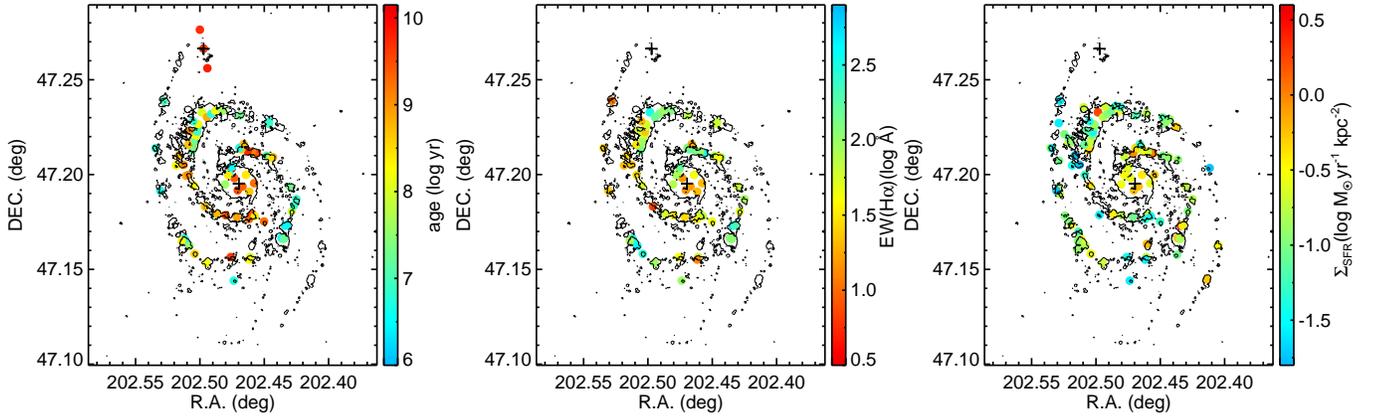}}}
\caption{Left: spatial distribution of light-weighted mean stellar population age in logarithmic scale. Middle: spatial distribution of EW({\Ha}). Right: spatial distribution of $\Sigma_{\mathrm{SFR}}$ in logarithmic scale. The contours display the isophotal shapes of the {\Ha} emission. The centers of M51 and NGC 5195 are marked with pluses.}
\label{fig:age}
\end{figure*}

The left panel of Figure \ref{fig:age} shows the distribution of the stellar age. The middle panel of Figure \ref{fig:age} shows the spatial distribution of the equivalent width of the {\Ha} line (EW({\Ha})). The EW({\Ha}) is sensitive to the ratio of present to past SFRs \citep[see ][ and references therein]{kong04} and is related to the strength of ionization of the ionization source \citep{lacerda18,sanchez19}. We have eliminated the regions influenced by AGN. The EW({\Ha}) of  the majority of our samples are larger than 14 {\AA}, so they are related to the ionization due to young stellar population. The maps of the stellar age and EW({\Ha}) are generally consistent as shown in Figure \ref{fig:age}. The bulge of M51 is older than the outskirt and the inner arms are older than the outer arms. This kind of decreasing radial age profile supports the ``inside-out" galaxy growth scenario, which is suitable for a majority of disk galaxies \citep{sanchezB14,sanchez19}.  There are considerable {\HII} regions in the outer arms with age of 50 -- 500 Myr.  The recent close encounter between M51 and NGC 5195 can trigger substantial star formation and form young stellar populations. We also obtain the light-weighted mean stellar age of three regions in NGC 5195, including one in the galactic core and two in the disk. All of them are very old stellar population and the average age is about 10 Gyr. Almost no recent star formation is found in this galaxy, since there is lack of any {\Ha} radiation. 

\citet{cal05} reported that M51 is a typical star-forming galaxy with the total SFR $\sim$ 3.4 M$_\sun$ yr$^{-1}$ and $\Sigma_{\mathrm{SFR}}$ = 0.015 M$_\sun$ yr$^{-1}$ kpc$^{-2}$. The right panel of Figure \ref{fig:age} shows $\Sigma_{\mathrm{SFR}}$ distribution of the {\HII} regions in M51. We can also see that there is on-going star formation in the bulge region of M51. There is no obvious difference of $\Sigma_{\mathrm{SFR}}$ between the northern and southern spiral arms. Although the close encounters can trigger star formation through the perturbation, they might disperse the gas more widely in the north, such that the north part presents similar $\Sigma_{\mathrm{SFR}}$ to the south. It can be also seen from the {\Ha} map that the {\Ha} emission looks more diffuse in the northern arms. Nevertheless, we need more observational evidence to support it.

\begin{figure}[!htb]
\center
\resizebox{\width}{!};{\rotatebox{-90}
{\includegraphics[width=0.45\textwidth]{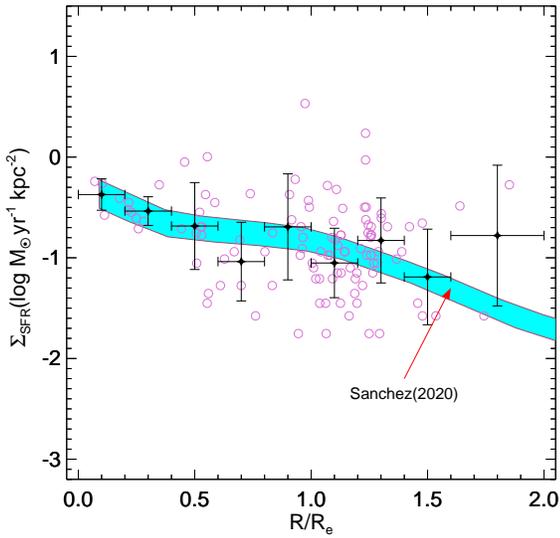}}}
\caption{Deprojected radial distribution of $\Sigma_{\mathrm{SFR}}$. The black star with error bar present the mean value and its standard deviation in each radial bin, while the horizontal bars show the range of each bin. The radial distribution of $\Sigma_{\mathrm{SFR}}$ for Sbc-type galaxies with $10^{10.5}$ M$_{\sun}$ $< M_* < 10^{11}$ M$_{\sun}$ from \citet{sanchez19} is shifted upward by 1.4 dex to match our data and overplotted in cyan.}
\label{fig:sfrgra}
\end{figure}

In Figure \ref{fig:sfrgra}, we present the radial distribution of $\Sigma_{\mathrm{SFR}}$. From this figure, we can see there is a monotonic decrease of the radial $\Sigma_{\mathrm{SFR}}$ in M51, although the radial profile presents a relatively large dispersion. The gradient is about -0.55 dex $R_e^{-1}$.  \citet{gonzalez16} characterized the radial structure of $\Sigma_{\mathrm{SFR}}$ of the CALIFA galaxies, and they found that $\Sigma_{\mathrm{SFR}}$ of all spiral galaxies decreases with radial distance. The typical gradient in the central $1 \times R_e$ is about -0.78 dex/$R_e$. \citet{sanchez19} presented the radial distributions of $\Sigma_{\mathrm{SFR}}$ for low-redshift galaxies of different stellar mass and morphology based on the recent IFS surveys.  M51 is a Sbc-type galaxy and its stellar mass is about 4.7$\times$10$^{10}$ M$_{\sun}$ \citep{men12}. We overlay the radial $\Sigma_{\mathrm{SFR}}$ profile from \citet{sanchez19} for galaxies with the same morphological type and similar stellar mass in Figure \ref{fig:sfrgra}. It shows that the radial $\Sigma_{\mathrm{SFR}}$ gradient of M51 is similar to the galaxies of similar mass and morphology in the nearby universe.

\subsection{Chemical Abundance Distribution} \label{subsec:metallicity}

\subsubsection{The Spatial Distribution of Oxygen Abundances} \label{subsubsec:distribution}
 
\begin{figure*}[!htb]
\center
\resizebox{\width}{!}{\rotatebox{-90}
{\includegraphics[width=0.5\textwidth]{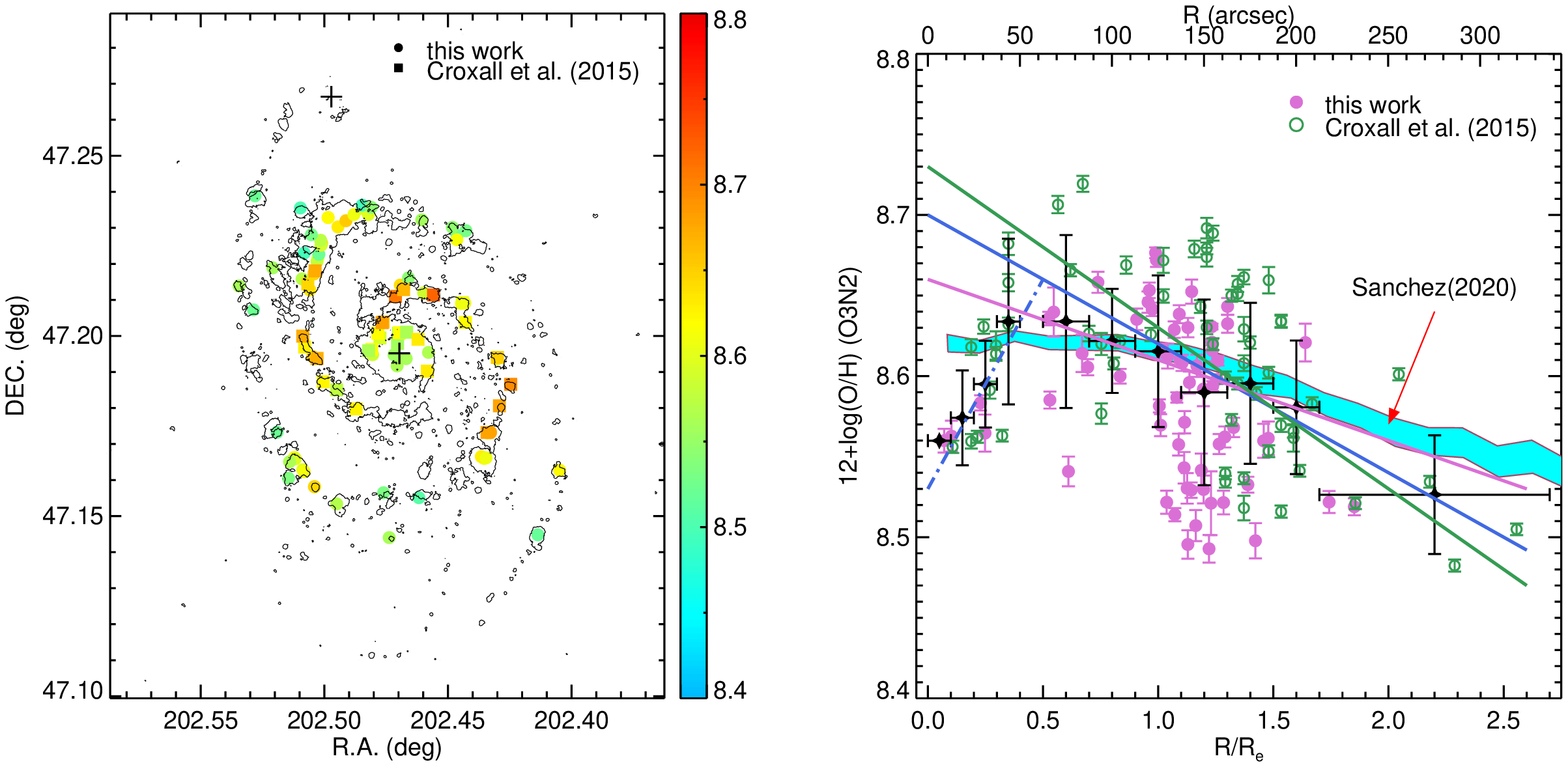}}}
\caption{Left: two-dimensional distribution of the oxygen abundance of {\HII} regions in M51. The contours display the isophotal shapes of the {\Ha} emission. The centers of M51 and NGC 5195 are marked with pluses. Right: radial distribution of the oxygen abundance. The filled circles and open circles are the data from this work and \citet{croxall15}, respectively. The black stars with error bars are the mean values, while the horizontal bars show the range of each bin. The blue dashed line shows the positive slop at $R/R_e < 0.4$, while the blue solid line shows the negative gradient at $R/R_e >0.4$. The violescent and green lines display the gradients for the data in our work and the data of \citet{croxall15} at $R/R_e >0.4$, respectively. The radial distribution of oxygen abundance for Sbc-type galaxies with $10^{10.5}$ M$_{\sun}$ $< M_* < 10^{11}$ M$_{\sun}$  from \citet{sanchez19} is shifted to match our data and overplotted in cyan.}
\label{fig:metallicity}
\end{figure*}

There are a total of 67 {\HII} regions in M51 whose gas-phase oxygen abundances are estimated using the O3N2 calibration. The two-dimensional distribution of the oxygen abundance is shown in the left panel of Figure \ref{fig:metallicity}. The data obtained by \citet{croxall15} are also overplotted in this figure. Although \citet{croxall15} derived the oxygen abundance through the direct $T_e$ method, we recalculate the metallicity using the same O3N2 calibration as used in this paper. Note that different strong-line calibrators may give distinct absolute values of metallicity.

The right panel of Figure \ref{fig:metallicity} shows the radial profile of the oxygen abundance. The data points from \citet{croxall15}, recalculated by with the same O3N2 calibration, are also presented in this figure. These two datasets are complementary. It is interesting that there are two different gradients in the inner and outer regions. The inner region presents a positive gradient and the outer disk region shows a negative one. We perform two independent linear fits to the radial profile. The cutoff point is manually set at $R/R_{e} = 0.4$, which is around the turn point of the radial profile as seen in Figure \ref{fig:metallicity}.  It is corresponding to the angular distance of about 50\arcsec and the galactocentric distance of 1.9 kpc. The region of $R/R_e < 0.4$ includes the bulge \citep[size of 11\arcsec$\times$16\arcsec, ][]{lamers02} and surrounding star forming ring. The linear fits are expressed as
\begin{eqnarray}
&12&+\log{(\mathrm{O/H})}[\mathrm{O3N2}] \nonumber \\
&=& (8.70\pm0.02) - (0.08\pm0.015) \times R/R_{e} \nonumber \\
&=& (8.70\pm0.02) - (0.22\pm0.04) \times R/R_{25}, \label{equ:grad1}
\end{eqnarray}
for $R/R_{e} > 0.4$ or $R/R_{25} > 0.15$, and
\begin{eqnarray}
&12&+\log{(\mathrm{O/H})}[\mathrm{O3N2}] \nonumber \\
&=& (8.53\pm 0.02) + (0.26\pm0.09) \times R/R_{e} \nonumber \\
&=& (8.53\pm 0.02) + (0.71\pm0.25) \times R/R_{25},  \label{equ:grad2}
\end{eqnarray}
for $0 < R/R_{e} < 0.4$ or $0 < R/R_{25} < 0.15$. In Figure \ref{fig:metallicity}, we also present the linear fits with only our data and only the data of \citet{croxall15}. The corresponding gradients in the outer region are $-0.05\pm0.025$ and $-0.10\pm0.02$ dex R$_{e}^{-1}$, respectively. These gradients are consistent with that of the combined data within the uncertainty. Combining two datasets should provide more reliable gradient measurements, because we use the same metallicity calibrator. The metallicity gradients as shown in Equation (\ref{equ:grad1}) and (\ref{equ:grad2}) are adopted in the following analyses.

\subsubsection{Comparison with Previous Work} \label{subsubsec:comparison}

The radial negative gradient of the oxygen abundance in the disk region (-0.22$\pm$0.04 dex R$_{25}^{-1}$) is close to other measurements based on different diagnostic methods. \citet{moustakas10} used the empirical $R_{23}$ calibration of \citet{pilyugin05} and provided an abundance gradient of about -0.31 $\pm$ 0.06 dex R$_{25}^{-1}$ for about 20 {\HII} regions. \citet{bre04} used a sample of 10 {\HII} regions with the $T_{e}$ measurements and obtained a gradient of -0.28 $\pm$ 0.14 dex R$_{25}^{-1}$. \citet{croxall15} derived an abundance gradient of -0.30 $\pm$ 0.10 dex R$_{25}^{-1}$ through the temperature-sensitive auroral lines for 29 individual {\HII} regions. Considering different calibration methods and limited {\HII} samples, our estimate of the metallicity gradient for the M51 disk is in agreement with the above measurements within the uncertainties. Our abundance gradient is statistically more accurate, since we include a large number of {\HII} regions. 

It is well known that negative radial gradients of the gas-phase metallicity are ubiquitous in many disk galaxies \citep{sear71,oey93,zari94,moustakas10}. It is can be simulated by chemical evolution models, indicating an ``inside-out" galaxy growth. The oxygen abundance gradient of the M51 disk (-0.22 $\pm$ 0.04 dex R$_{25}^{-1}$) is shallower than some other isolated spiral galaxies, such as NGC 628 \citep[-0.485 $\pm$ 0.122 dex R$_{25}^{-1}$,][]{berg15}, NGC 2403 \citep[-0.524 $\pm$ 0.043 dex R$_{25}^{-1}$, ][]{pilyugin14}, and M101 \citep[-0.832 $\pm$ 0.044 dex R$_{25}^{-1}$, ][]{croxall16}. However, such a comparison is less persuasive due to a small sample. \citet{kewley10} and \citet{sanchez14} found evidence that the interacting systems present shallower metallicity gradients compared to the isolated galaxies. Recently, the IFS surveys have offered the opportunity to achieve meaningful statistical results about abundance gradients for large samples of galaxies in the Local universe with different properties, such as morphology, stellar mass, and environment density  \citep{sanchez14, ho15, belfiore17, sanchezM18, sanchez19}.  Using the high spatial resolution IFS data obtained by MUSE, \citet{sanchezM18} found that spiral galaxies present a characteristic abundance slope of $-0.10 \pm 0.03 $ dex/$R_e$ between 0.5 $R_e$ and 1.5 $R_e$.  Our gradient measurement of the M51 disk is consistent with this characteristic slope. \citet{sanchezM18} also studied the possible effect of the density of the galaxy environment on the metallicity gradient and claimed that their spiral galaxy samples present a similar slope independent of the environment. It is worthy to notice that they discarded those distorted galaxies with recent interactions (like the M51-NGC 5195 system) in their analysis. \citet{sanchez19} showed the radial profiles of the oxygen abundance for galaxies of different morphology and stellar mass. From Figure 15 in \citet{sanchez19}, the typical gradient is about -0.06 dex $R_e^{-1}$ for $10^{10.5}$ M$_{\sun}$ $< M_* < 10^{11}$ M$_{\sun}$ and Sbc type in the range of $0.4 R_e < R < 2 R_e$. We also present the radial metallicity profile from \citet{sanchez19} in the Figure \ref{fig:metallicity}. It seems that the M51 disk does not present much difference of the metallicity gradient from those galaxies with similar stellar mass and morphological type within the measurement uncertainty. According to the above comparison with the literature, the interaction does not seem to have a significant impact in the metallicity gradient measured in the disk of M51.

From the right panel of Figure \ref{fig:metallicity}, we can see that a dozen {\HII} regions at $R < 0.4 R_e$ show systematically lower metallicities, compared to the extrapolated oxygen abundance of the outer disk into the inner region. There have been observational evidences based on IFU data that some galaxies present some different behaviors from a simple negative radial gradient in the innermost and outer parts \citep{sanchez14, sanchezM18, sanchez19}. These studies found that the oxygen abundance of galaxies sometimes presents a prominent flattening or drop in the inner regions. The inner drop is ubiquitous in massive galaxies with $M_{\ast} > 10^{10} M_{\sun}$ \citep[][ and references therein]{sanchez19} and it always appears at a similar location of $R \sim 0.5 R_e$ for all galaxies \citep{sanchezM18}. M51 presents a drop at almost the same location. A possible cause to the inner drop as mentioned by \citet{sanchez19} is the effect of radial migration to the Lindblad resonances or the freezing of the chemical enrichment due to the star-formation quenching in the bulge-dominated inner region. \citet{sanchez15} claimed that the ionization conditions of {\HII} regions seem to be closely related to the properties of the underlying stellar population. It means that the metal enrichment of an {\HII} region is not necessarily correlated with the most recent star formation but rather to the local star-formation history.  Recent encounters in M51 may induce gas infall and form new {\HII} regions in the inner region, but they might keep a memory of the local metal enrichment in the history, when the chemical enrichment could be frozen due to the star-formation quenching. Thus, the inner region of M51 might still keep the relative poor metallicity and cause a positive metallicity gradient.

On the other hand, we also notice that the bulge of M51 is a so-called pseudobulge \citep{fis10}, which presents a near-exponential surface brightness profile, rotating motion, active star formation, nuclear bar, ring and/or spiral \citep{Kormendy04}. The non-axisymmetric nuclear bar and spiral arms can drive the gas infall, facilitate the pseudobulge growth, and cause higher gas-phase extinction and active star formation in the inner region of M51 as shown in Figure \ref{fig:av} and \ref{fig:age}. A rather chaotic distribution of dust lanes can be found in high-resolution images of the central region of M51 taken by the Hubble Space Telescope, which also indicates that M51 is transporting gas to the nucleus \citep{grillmair97}. The gas inflow into the galactic center might dilute the metallicity and hence form a positive gradient. Another possible reason for the inner drop is the interaction between M51 and NGC 5195, which might also induce the gas inflow and the radial mixing.

\section{Summary} \label{sec:summary}
Nearby galaxies are ideal laboratories for understanding the galaxy formation and evolution in great details. M51 is undergoing the gravitational interaction with its companion of NGC 5195 at the distance of about 7.9 Mpc. It is an excellent object to investigate the spatial distributions of the physical properties and seek the possible clues of the formation and evolution for similar galaxies. There are substantial {\HII} regions across the whole galaxy of M51. As one of 20 large nearby spiral galaxies in our observing program \citep{kong14}, a large number of {\HII} regions of M51 have been observed by using the NAOC 2.16 m and 6.5 m MMT telescopes. A total of 113 spectra are obtained:  99 of them are from the long-slit spectrograph and the rest are the fiber spectra. Together with the literature data of \citet{croxall15}, we derive the two-dimensional distributions and corresponding radial gradients of a series of physical properties and try to explore the possible evolution clues of this galaxy and the influence of the galactic interaction. Through the emission line measurements and spectral fitting, we obtain the gas-phase extinction, SFR surface density, stellar population age,  oxygen abundance, and their corresponding spatial distributions. Some of the main points in this paper is summarized as follows:
\begin{enumerate}[(1)]
\item There is a mild radial extinction gradient. The gas-phase extinction in the northern spiral arms that are close to NGC 5195 is lower than that of the southern arms. It might be related to the galactic interaction, which disperses the gas distribution in the area close to the companion. 
\item M51 has a number of young {\HII} regions with age of 50--500 Myr, which is consistent with the recent close interaction history with NGC 5195. Similar to most spiral galaxies, M51 presents a mildly radial gradient of the SFR surface density.
\item Three spectra for NGC 5195 are obtained in its bulge and disk. It is presented that NGC 5195 is an old galaxy with average age of about 10 Gyr and no recent star formation occurs in this galaxy.
\item There is a radial negative gradient of gas-phase metallicity in the disk of M51 (-0.08 dex $R_e^{-1}$). The age distribution and the radial negative gradient supports the ``inside-out" galaxy growth. There is no clear evidence that the metallicity gradient is flattened by the galactic interaction. 
\item There is a positive metallicity slope in the inner region, which include the bulge and surrounding star forming ring. This metallicity drop might be caused by the freezing of the chemical enrichment due to the star-forming quenching in the bulge-dominated inner region. Another possible reason is the growth of the pseudobulge and/or galactic interaction, which might induce the gas infall and dilute the metallicity in the center. 
\end{enumerate}

\acknowledgments
We thank the anonymous referee for his/her thoughtful comments and insightful suggestions that improve our paper greatly. This work is supported by the Xinjiang Natural Science Foundation (No. 2020D01B59) and Major Program of National Natural Science Foundation of China (No. 11890691). This work is supported by the National Key R\&D Program of China (973 Program; grant Nos. 2017YFA0402600), the National Natural Science Foundation of China (NSFC, grant Nos. 11673027, 11733007, 11973038, 11320101002, 11421303, 11890693), and the External Cooperation Program of Chinese Academy of Sciences (grant No. 114A11KYSB20160057). This work uses the observational time of the 2.16m telescope at the Xinglong station of the National Astronomical Observatories of China and the observational time of the MMT telescope obtained via the Telescope Access Program (TAP), which is funded by the National Astronomical Observatories of China, the Chinese Academy of Sciences (the Strategic Priority Research Program, “The Emergence of Cosmological Structures” grant No. XDB09000000), and the Special Fund for Astronomy from the Ministry of Finance. The 2.16m telescope is jointly operated and administrated by the National Astronomical Observatories of China and Center for Astronomical Mega-Science, Chinese Academy of Sciences.

%






\appendix
\section{Physical Properties of {\HII} regions in M51} \label{appe:tab2} 
We present the physical properties derived from the observed spectra in this paper, which is shown in Table \ref{tab:flux}. This table contains the emission line measurements and corresponding derivatives. For more information, we can refer to Section \ref{sec:measurement}.

\begin{longrotatetable}
\begin{deluxetable*}{lcccccccccccc}
\tablecaption{Emission-line Measurements and Spectral Properties\label{tab:flux}}
\tablewidth{4000pt}
\tabletypesize{\scriptsize}
\tablehead{\colhead{ID}  & \colhead{R.A.} & \colhead{Decl.}  & \colhead{$R/R_{25}$} & \colhead{[{\OIII}]} & \colhead{{\Ha}} & \colhead{[{\NII}]} & \colhead{{\Hb}} & \colhead{EW({\Ha})} & \colhead{$E(B-V)$} &  \colhead{12+log(O/H)} & 
\colhead{$\Sigma_{\mathrm{SFR}}$} & \colhead{$\log \mathrm{(age)}$} \\
\colhead{(1)}  & \colhead{(2)}  & \colhead{(3)}  & \colhead{(4)} & \colhead{(5)} & \colhead{(6)} & \colhead{(7)} & \colhead{(8)} & \colhead{(9)} & \colhead{(10)} & \colhead{(11)} & \colhead{(12)} & \colhead{(13)} } 
\startdata
  1  &     13:30:1.318 &  47:12:49.427  &    0.341   &             &     2.942   &     1.087 &    1553.557 &   89.963  &   0.434 &           &   0.238  &   8.887  \\
     &                 &                &            &             &     0.017   &     0.008 &      24.614 &    0.513  &   0.016 &           &          &          \\
  2  &     13:30:0.974 &   47:13:3.463  &    0.357   &     0.086   &     2.885   &     0.900 &     823.109 &   80.628  &   0.134 &    8.653  &   0.124  &   8.450  \\
     &                 &                &            &     0.004   &     0.008   &     0.004 &       9.541 &    0.217  &   0.011 &    0.005  &          &          \\
  3  &     13:30:0.626 &  47:13:17.758  &    0.379   &     0.146   &     2.874   &     1.021 &     433.346 &   51.129  &   0.075 &    8.616  &   0.065  &   8.359  \\
     &                 &                &            &     0.006   &     0.016   &     0.009 &       6.993 &    0.278  &   0.016 &    0.004  &          &          \\
  4  &     13:30:0.260 &  47:13:32.741  &    0.406   &     0.108   &     2.871   &     0.967 &     390.016 &   54.416  &   0.058 &    8.638  &   0.058  &   7.398  \\
     &                 &                &            &     0.006   &     0.007   &     0.004 &       4.418 &    0.129  &   0.011 &    0.006  &          &          \\
  5  &    13:29:59.634 &  47:13:58.326  &    0.458   &     0.139   &     2.913   &     1.159 &    6142.437 &  253.225  &   0.282 &    8.631  &   0.933  &   7.110  \\
     &                 &                &            &     0.002   &     0.009   &     0.004 &      68.908 &    0.785  &   0.011 &    0.002  &          &          \\
  6  &     13:30:1.403 &  47:12:45.939  &    0.337   &     0.126   &     3.010   &     1.141 &    2701.989 &  136.221  &   0.787 &    8.635  &   0.424  &   7.245  \\
     &                 &                &            &     0.009   &     0.023   &     0.009 &      58.961 &    1.056  &   0.022 &    0.007  &          &          \\
  7  &     13:30:1.014 &   47:13:1.924  &    0.355   &     0.089   &     2.921   &     0.871 &    1335.628 &   98.341  &   0.327 &    8.646  &   0.204  &   7.205  \\
     &                 &                &            &     0.006   &     0.011   &     0.004 &      22.093 &    0.379  &   0.016 &    0.007  &          &          \\
  8  &     13:30:0.707 &  47:13:14.408  &    0.374   &     0.217   &     2.912   &     1.063 &     603.281 &   61.893  &   0.278 &    8.582  &   0.092  &   8.330  \\
     &                 &                &            &     0.008   &     0.016   &     0.008 &      15.570 &    0.344  &   0.025 &    0.004  &          &          \\
  9  &     13:30:0.531 &  47:13:21.535  &    0.385   &     0.472   &     2.914   &     1.217 &     245.681 &   22.338  &   0.287 &    8.522  &   0.037  &   6.913  \\
     &                 &                &            &     0.024   &     0.033   &     0.022 &      12.876 &    0.252  &   0.051 &    0.007  &          &          \\
 10  &     13:30:0.311 &  47:13:30.654  &    0.402   &     0.197   &     2.905   &     1.021 &     687.594 &   62.071  &   0.242 &    8.587  &   0.104  &   6.976  \\
     &                 &                &            &     0.006   &     0.012   &     0.006 &       9.451 &    0.262  &   0.014 &    0.003  &          &          \\
 11  &    13:29:59.634 &  47:13:58.326  &    0.458   &     0.143   &     2.944   &     1.084 &   11208.229 &  329.473  &   0.449 &    8.621  &   1.721  &   8.005  \\
     &                 &                &            &     0.002   &     0.019   &     0.007 &     141.557 &    2.130  &   0.014 &    0.002  &          &          \\
 12  &    13:29:44.879 &  47:09:27.650  &    0.471   &             &     2.967   &     1.117 &     550.305 &   82.368  &   0.569 &           &   0.085  &          \\    
     &                 &                &            &             &     0.036   &     0.017 &      36.247 &    0.987  &   0.064 &           &          &          \\
 13  &    13:29:44.491 &  47:09:57.670  &    0.407   &     0.131   &     2.908   &     0.873 &    1489.537 &  126.645  &   0.256 &    8.610  &   0.226  &   7.772  \\
     &                 &                &            &     0.002   &     0.007   &     0.003 &      14.316 &    0.308  &   0.010 &    0.002  &          &          \\
 14  &    13:29:44.165 &  47:10:23.048  &    0.362   &     0.107   &     2.940   &     1.012 &   22199.240 &  671.499  &   0.427 &    8.641  &   3.405  &   6.871  \\
     &                 &                &            &     0.004   &     0.031   &     0.009 &     446.091 &    7.158  &   0.022 &    0.004  &          &          \\
 15  &    13:29:43.206 &  47:11:37.137  &    0.310   &     0.154   &     2.913   &     0.918 &    1177.172 &   63.185  &   0.281 &    8.600  &   0.179  &   7.281  \\
     &                 &                &            &     0.007   &     0.010   &     0.005 &      22.354 &    0.211  &   0.018 &    0.005  &          &          \\
 16  &     13:30:2.113 &  47:11:59.316  &    0.310   &             &     2.921   &     0.959 &     368.112 &   28.493  &   0.326 &           &   0.056  &   9.025  \\
     &                 &                &            &             &     0.013   &     0.008 &       8.436 &    0.130  &   0.022 &           &          &          \\
 17  &     13:30:3.047 &  47:12:18.364  &    0.351   &             &     2.904   &     0.892 &     131.759 &   23.515  &   0.233 &           &   0.020  &   9.029  \\
     &                 &                &            &             &     0.013   &     0.008 &       3.151 &    0.103  &   0.023 &           &          &          \\
 18  &     13:30:3.918 &  47:12:36.216  &    0.396   &             &     2.898   &     1.024 &     137.162 &   27.585  &   0.201 &           &   0.021  &   8.755  \\
     &                 &                &            &             &     0.011   &     0.007 &       2.692 &    0.109  &   0.019 &           &          &          \\
 19  &     13:30:5.127 &   47:13:0.949  &    0.463   &             &     2.962   &     0.903 &    1366.686 &  140.693  &   0.543 &           &   0.211  &          \\    
     &                 &                &            &             &     0.027   &     0.010 &      55.393 &    1.267  &   0.039 &           &          &          \\
 20  &     13:30:5.896 &  47:13:16.646  &    0.508   &             &     2.926   &     0.868 &     639.684 &  129.887  &   0.351 &           &   0.098  &          \\    
     &                 &                &            &             &     0.010   &     0.004 &      41.428 &    0.452  &   0.062 &           &          &          \\
 21  &     13:30:6.932 &  47:13:37.850  &    0.571   &             &     2.925   &     1.192 &     153.572 &   63.189  &   0.347 &           &   0.023  &          \\    
     &                 &                &            &             &     0.049   &     0.028 &       8.675 &    1.061  &   0.056 &           &          &          \\
 22  &    13:29:44.689 &  47:12:28.622  &    0.298   &             &     2.962   &     1.155 &     720.355 &   93.450  &   0.540 &           &   0.111  &          \\    
     &                 &                &            &             &     0.056   &     0.026 &      32.720 &    1.779  &   0.047 &           &          &          \\
 23  &    13:29:46.362 &  47:12:32.866  &    0.258   &     0.155   &     2.906   &     0.983 &    1004.989 &   98.683  &   0.244 &    8.606  &   0.152  &   8.869  \\
     &                 &                &            &     0.008   &     0.015   &     0.006 &      15.312 &    0.498  &   0.015 &    0.005  &          &          \\
 24  &    13:29:47.560 &  47:12:35.887  &    0.233   &             &     2.903   &     0.704 &     638.925 &   68.805  &   0.230 &           &   0.097  &   7.079  \\
     &                 &                &            &             &     0.026   &     0.011 &      15.108 &    0.606  &   0.024 &           &          &          \\
 25  &    13:29:49.347 &  47:12:40.432  &    0.206   &     0.001   &     2.986   &     1.080 &    6499.730 &  307.407  &   0.664 &           &   1.012  &   9.420  \\
     &                 &                &            &             &     0.028   &     0.010 &     279.806 &    2.867  &   0.042 &           &          &          \\
 26  &    13:29:50.354 &  47:12:43.014  &    0.197   &     0.139   &     2.934   &     0.713 &    1322.978 &   94.027  &   0.392 &    8.585  &   0.202  &   9.105  \\
     &                 &                &            &     0.007   &     0.035   &     0.015 &      27.473 &    1.125  &   0.023 &    0.005  &          &          \\
 27  &    13:29:51.083 &  47:12:44.854  &    0.194   &             &     2.965   &     1.023 &    1841.999 &  123.719  &   0.557 &           &   0.285  &   9.713  \\
     &                 &                &            &             &     0.026   &     0.011 &      45.989 &    1.068  &   0.025 &           &          &          \\
 28  &    13:29:52.295 &  47:12:47.917  &    0.195   &     0.092   &     2.904   &     0.806 &    1350.585 &  177.085  &   0.236 &    8.636  &   0.205  &   8.667  \\
     &                 &                &            &     0.004   &     0.016   &     0.006 &      17.255 &    0.983  &   0.013 &    0.004  &          &          \\
 29  &    13:29:45.026 &  47:13:30.791  &    0.412   &             &     2.977   &     1.189 &    3051.277 &  266.865  &   0.617 &           &   0.474  &          \\    
     &                 &                &            &             &     0.054   &     0.021 &     157.654 &    4.818  &   0.052 &           &          &          \\
 30  &    13:29:47.109 &  47:13:40.363  &    0.399   &     0.453   &     2.889   &     1.067 &     690.511 &   76.383  &   0.154 &    8.514  &   0.104  &   6.521  \\
     &                 &                &            &     0.016   &     0.025   &     0.011 &      16.485 &    0.656  &   0.024 &    0.004  &          &          \\
 31  &    13:29:50.629 &  47:09:23.297  &    0.419   &             &     2.983   &     1.020 &    1064.178 &   61.779  &   0.648 &           &   0.166  &   7.941  \\
     &                 &                &            &             &     0.027   &     0.013 &      42.680 &    0.551  &   0.039 &           &          &          \\
 32  &    13:29:50.500 &  47:09:48.318  &    0.346   &             &     2.987   &     0.795 &    3885.289 &  306.589  &   0.669 &           &   0.605  &          \\    
     &                 &                &            &             &     0.032   &     0.009 &     148.490 &    3.306  &   0.038 &           &          &          \\
 33  &    13:29:50.211 &  47:10:45.049  &    0.188   &             &     2.953   &     0.466 &    1136.890 &   67.431  &   0.491 &           &   0.175  &   9.460  \\
     &                 &                &            &             &     0.030   &     0.010 &      75.436 &    0.680  &   0.064 &           &          &          \\
 34  &    13:29:44.762 &  47:09:59.469  &    0.398   &     0.110   &     2.936   &     0.910 &    3401.850 &  102.866  &   0.403 &    8.629  &   0.521  &   6.541  \\
     &                 &                &            &     0.006   &     0.013   &     0.005 &      41.733 &    0.468  &   0.012 &    0.005  &          &          \\
 35  &    13:29:43.868 &  47:10:23.378  &    0.369   &     0.080   &     2.918   &     1.035 &    2115.448 &  206.212  &   0.307 &    8.672  &   0.322  &   6.406  \\
     &                 &                &            &     0.003   &     0.027   &     0.010 &      41.355 &    1.895  &   0.021 &    0.004  &          &          \\
 36  &    13:29:42.858 &  47:10:50.267  &    0.354   &             &     2.934   &     0.799 &     664.662 &   52.532  &   0.394 &           &   0.102  &   7.867  \\
     &                 &                &            &             &     0.033   &     0.015 &      24.363 &    0.592  &   0.036 &           &          &          \\
 37  &    13:29:41.993 &  47:11:13.421  &    0.359   &             &     2.971   &     0.925 &    1043.371 &   85.908  &   0.589 &           &   0.162  &   6.888  \\
     &                 &                &            &             &     0.031   &     0.013 &      33.942 &    0.904  &   0.032 &           &          &          \\
 38  &    13:29:39.272 &  47:08:41.315  &    0.688   &     0.439   &     2.953   &     1.110 &    3466.517 &  487.261  &   0.492 &    8.519  &   0.534  &          \\    
     &                 &                &            &     0.013   &     0.031   &     0.010 &     161.828 &    5.145  &   0.045 &    0.005  &          &          \\
 39  &    13:29:37.156 &  47:09:45.832  &    0.609   &     0.164   &     2.967   &     1.250 &    2137.881 &  286.519  &   0.567 &    8.621  &   0.331  &          \\    
     &                 &                &            &     0.015   &     0.062   &     0.024 &     179.446 &    6.035  &   0.082 &    0.012  &          &          \\
 40  &    13:29:43.876 &  47:10:23.982  &    0.367   &     0.072   &     2.924   &     0.988 &    2434.352 &  230.308  &   0.342 &    8.677  &   0.371  &   6.497  \\
     &                 &                &            &     0.002   &     0.021   &     0.008 &      35.717 &    1.639  &   0.015 &    0.003  &          &          \\
 41  &    13:29:48.025 &  47:10:30.862  &    0.260   &             &     2.936   &     1.035 &     349.008 &   41.773  &   0.407 &           &   0.053  &   9.264  \\
     &                 &                &            &             &     0.026   &     0.015 &      21.298 &    0.365  &   0.058 &           &          &          \\
 42  &    13:29:50.826 &  47:10:35.504  &    0.208   &             &     2.919   &     0.836 &     284.159 &   35.640  &   0.316 &           &   0.043  &   9.326  \\
     &                 &                &            &             &     0.021   &     0.013 &      14.807 &    0.259  &   0.050 &           &          &          \\
 43  &    13:29:53.108 &  47:10:39.322  &    0.189   &             &     2.950   &     0.889 &     586.697 &   27.272  &   0.476 &           &   0.090  &   8.868  \\
     &                 &                &            &             &     0.011   &     0.008 &      17.782 &    0.101  &   0.029 &           &          &          \\
 44  &    13:29:55.184 &  47:10:42.741  &    0.197   &             &     2.992   &     1.609 &    1053.441 &   33.853  &   0.697 &           &   0.164  &   9.387  \\
     &                 &                &            &             &     0.021   &     0.015 &      71.388 &    0.239  &   0.065 &           &          &          \\
 45  &    13:29:56.781 &  47:10:45.406  &    0.218   &             &     3.008   &     0.924 &    2256.569 &   58.309  &   0.776 &           &   0.354  &   8.706  \\
     &                 &                &            &             &     0.011   &     0.006 &      57.247 &    0.208  &   0.024 &           &          &          \\
 46  &     13:30:0.959 &  47:09:29.408  &    0.484   &     0.148   &     2.881   &     1.247 &    1630.106 &  279.099  &   0.112 &    8.632  &   0.245  &   8.529  \\
     &                 &                &            &     0.007   &     0.022   &     0.009 &      60.678 &    2.171  &   0.036 &    0.006  &          &          \\
 47  &     13:30:2.091 &  47:09:46.368  &    0.467   &     0.186   &     2.914   &     1.092 &    1106.449 &  332.863  &   0.286 &    8.598  &   0.168  &   6.538  \\
     &                 &                &            &     0.004   &     0.017   &     0.006 &      24.232 &    1.894  &   0.022 &    0.003  &          &          \\
 48  &     13:30:2.900 &  47:09:58.576  &    0.459   &     0.158   &     2.914   &     1.118 &    1573.680 &  244.229  &   0.289 &    8.616  &   0.239  &   6.306  \\
     &                 &                &            &     0.011   &     0.020   &     0.007 &      40.949 &    1.675  &   0.025 &    0.007  &          &          \\
 49  &     13:30:4.585 &  47:10:23.804  &    0.457   &     0.443   &     2.979   &     1.165 &    2033.249 &  248.794  &   0.631 &    8.521  &   0.316  &          \\    
     &                 &                &            &     0.077   &     0.103   &     0.039 &     220.600 &    8.594  &   0.108 &    0.020  &          &          \\
 50  &     13:30:0.959 &  47:09:29.408  &    0.484   &     0.140   &     2.761   &     1.268 &    1895.726 &  481.286  &         &    8.643  &   0.273  &          \\    
     &                 &                &            &     0.006   &     0.019   &     0.007 &      76.012 &    3.320  &   0.039 &    0.005  &          &          \\
 51  &     13:30:1.956 &  47:09:44.390  &    0.468   &     0.185   &     2.907   &     1.225 &    1113.187 &  406.040  &   0.254 &    8.609  &   0.169  &   7.594  \\
     &                 &                &            &     0.003   &     0.024   &     0.009 &      29.632 &    3.375  &   0.026 &    0.003  &          &          \\
 52  &     13:30:2.812 &  47:09:57.244  &    0.460   &     0.227   &     2.890   &     1.264 &    2136.803 &  292.724  &   0.162 &    8.594  &   0.322  &   6.823  \\
     &                 &                &            &     0.005   &     0.021   &     0.008 &      48.534 &    2.092  &   0.023 &    0.003  &          &          \\
 53  &     13:30:5.116 &  47:10:31.796  &    0.460   &             &     2.923   &     1.322 &     692.242 &  256.869  &   0.338 &           &   0.106  &          \\    
     &                 &                &            &             &     0.078   &     0.035 &      71.218 &    6.887  &   0.101 &           &          &          \\
 54  &     13:30:5.804 &  47:10:42.096  &    0.467   &             &     2.939   &     0.793 &     684.261 &  117.833  &   0.421 &           &   0.105  &          \\    
     &                 &                &            &             &     0.063   &     0.027 &      49.141 &    2.513  &   0.071 &           &          &          \\
 55  &     13:30:3.475 &  47:09:37.840  &    0.516   &     0.408   &     2.917   &     1.188 &     750.475 &  184.806  &   0.303 &    8.533  &   0.114  &          \\    
     &                 &                &            &     0.018   &     0.027   &     0.011 &      19.527 &    1.680  &   0.026 &    0.005  &          &          \\
 56  &     13:30:3.376 &  47:09:54.292  &    0.479   &     0.265   &     2.946   &     1.068 &     752.139 &   90.336  &   0.459 &    8.562  &   0.116  &   8.304  \\
     &                 &                &            &     0.016   &     0.029   &     0.013 &      26.724 &    0.875  &   0.035 &    0.007  &          &          \\
 57  &     13:30:3.274 &  47:10:11.815  &    0.442   &             &     2.943   &     1.026 &     416.521 &   98.427  &   0.441 &           &   0.064  &  10.003  \\
     &                 &                &            &             &     0.034   &     0.014 &      40.861 &    1.141  &   0.094 &           &          &          \\
 58  &     13:30:2.322 &   47:14:7.911  &    0.529   &     0.636   &     2.933   &     1.283 &    1304.002 &  145.787  &   0.389 &    8.498  &   0.200  &          \\    
     &                 &                &            &     0.033   &     0.058   &     0.027 &     132.318 &    2.891  &   0.098 &    0.011  &          &          \\
 59  &    13:29:52.936 &  47:11:30.079  &    0.038   &     0.226   &     3.031   &     0.938 &    3476.111 &   15.156  &   0.894 &    8.563  &   0.550  &   9.455  \\
     &                 &                &            &     0.019   &     0.020   &     0.016 &     197.418 &    0.101  &   0.054 &    0.010  &          &          \\
 60$^{a}$  &    13:29:53.236 &  47:11:41.106  &    0.018   &     1.807   &     3.026   &     6.841 &    2585.291 &    5.713  &   0.868 &    8.554  &   0.408  &   9.676  \\
     &                 &                &            &     0.045   &     0.031   &     0.038 &     200.606 &    0.058  &   0.074 &    0.008  &          &          \\
 61  &    13:29:53.569 &  47:11:53.274  &    0.042   &             &     2.988   &     1.185 &    1711.401 &   15.184  &   0.674 &           &   0.267  &   9.376  \\
     &                 &                &            &             &     0.015   &     0.011 &      87.832 &    0.076  &   0.049 &           &          &          \\
 62  &    13:29:54.064 &  47:12:11.552  &    0.096   &             &     2.903   &     0.623 &    1312.535 &   81.076  &   0.229 &           &   0.199  &   9.870  \\
     &                 &                &            &             &     0.027   &     0.010 &      44.247 &    0.766  &   0.033 &           &          &          \\
 63  &    13:29:54.675 &  47:12:34.019  &    0.165   &             &     2.965   &     0.598 &    1249.179 &   65.953  &   0.554 &           &   0.193  &          \\    
     &                 &                &            &             &     0.049   &     0.019 &      88.499 &    1.082  &   0.069 &           &          &          \\
 64  &     13:30:6.746 &  47:11:23.171  &    0.463   &             &     2.880   &     1.024 &    1203.968 &  263.298  &   0.109 &           &   0.181  &          \\    
     &                 &                &            &             &     0.282   &     0.103 &     270.323 &   25.733  &   0.232 &           &          &          \\
 65  &     13:30:6.987 &  47:11:35.297  &    0.467   &             &     2.897   &     0.932 &    1031.370 &  166.016  &   0.197 &           &   0.156  &          \\    
     &                 &                &            &             &     0.060   &     0.021 &      65.357 &    3.442  &   0.063 &           &          &          \\
 66  &     13:30:8.492 &  47:12:50.718  &    0.549   &             &     2.940   &     0.862 &    1424.019 &  133.896  &   0.428 &           &   0.218  &          \\    
     &                 &                &            &             &     0.035   &     0.013 &     136.055 &    1.578  &   0.091 &           &          &          \\
 67  &    13:29:58.802 &  47:09:12.242  &    0.493   &     0.261   &     2.958   &     1.125 &    1310.088 &   92.598  &   0.519 &    8.568  &   0.202  &   8.187  \\
     &                 &                &            &     0.012   &     0.022   &     0.010 &      56.322 &    0.702  &   0.041 &    0.006  &          &          \\
 68  &    13:29:54.250 &  47:09:23.365  &    0.418   &     0.590   &     2.949   &     1.693 &     235.730 &   38.776  &   0.472 &    8.530  &   0.036  &   9.609  \\
     &                 &                &            &     0.045   &     0.041   &     0.026 &      15.722 &    0.534  &   0.065 &    0.009  &          &          \\
 69  &    13:29:49.761 &  47:10:40.805  &    0.206   &             &     2.874   &     0.661 &     228.848 &   65.277  &   0.073 &           &   0.034  &   8.434  \\
     &                 &                &            &             &     0.025   &     0.010 &       9.173 &    0.566  &   0.039 &           &          &          \\
 70  &     13:30:6.877 &  47:12:26.095  &    0.477   &     0.466   &     2.925   &     1.208 &     551.844 &   94.427  &   0.344 &    8.521  &   0.084  &          \\    
     &                 &                &            &     0.023   &     0.029   &     0.014 &      34.419 &    0.925  &   0.060 &    0.007  &          &          \\
 71  &     13:30:5.002 &   47:13:8.036  &    0.470   &     0.373   &     2.923   &     1.420 &     453.477 &   65.801  &   0.337 &    8.558  &   0.069  &          \\    
     &                 &                &            &     0.017   &     0.012   &     0.007 &      21.479 &    0.268  &   0.045 &    0.006  &          &          \\
 72  &     13:30:4.684 &   47:13:1.087  &    0.451   &             &     2.942   &     1.409 &     796.267 &   70.970  &   0.437 &           &   0.122  &          \\    
     &                 &                &            &             &     0.047   &     0.025 &      52.773 &    1.128  &   0.065 &           &          &          \\
 73  &     13:30:1.959 &  47:13:22.798  &    0.420   &     0.665   &     2.915   &     1.294 &     449.785 &   58.531  &   0.292 &    8.495  &   0.068  &   9.072  \\
     &                 &                &            &     0.035   &     0.054   &     0.031 &      32.993 &    1.086  &   0.072 &    0.009  &          &          \\
 74  &     13:30:0.355 &  47:13:35.584  &    0.414   &     0.263   &     2.927   &     1.163 &     872.811 &   57.987  &   0.359 &    8.571  &   0.133  &   8.307  \\
     &                 &                &            &     0.018   &     0.017   &     0.009 &      28.582 &    0.346  &   0.032 &    0.007  &          &          \\
 75  &    13:29:58.678 &  47:13:48.932  &    0.420   &     0.142   &     2.896   &     1.170 &    1094.139 &  143.473  &   0.193 &    8.630  &   0.165  &   8.806  \\
     &                 &                &            &     0.008   &     0.022   &     0.009 &      28.989 &    1.099  &   0.026 &    0.006  &          &          \\
 76  &    13:29:57.902 &  47:13:55.098  &    0.426   &     0.113   &     2.894   &     1.180 &     748.891 &  133.163  &   0.181 &    8.652  &   0.113  &   6.828  \\
     &                 &                &            &     0.008   &     0.034   &     0.015 &      24.864 &    1.574  &   0.033 &    0.007  &          &          \\
 77  &    13:29:57.129 &   47:14:1.251  &    0.434   &     0.194   &     2.905   &     1.235 &     548.942 &   91.334  &   0.239 &    8.606  &   0.083  &   8.134  \\
     &                 &                &            &     0.011   &     0.037   &     0.018 &      23.734 &    1.170  &   0.043 &    0.007  &          &          \\
 78  &    13:29:56.356 &   47:14:7.444  &    0.445   &     0.223   &     2.897   &     1.219 &     492.534 &  113.638  &   0.199 &    8.592  &   0.074  &          \\    
     &                 &                &            &     0.019   &     0.039   &     0.018 &      34.399 &    1.538  &   0.068 &    0.011  &          &          \\
 79  &    13:29:51.416 &  47:11:59.316  &    0.066   &             &     2.987   &     0.633 &    2510.877 &   43.190  &   0.671 &           &   0.391  &   8.381  \\
     &                 &                &            &             &     0.026   &     0.012 &     140.849 &    0.370  &   0.054 &           &          &          \\
 80  &    13:29:55.312 &  47:11:41.285  &    0.085   &     0.126   &     2.939   &     0.638 &    1873.356 &   75.878  &   0.419 &    8.583  &   0.287  &   7.608  \\
     &                 &                &            &     0.006   &     0.013   &     0.005 &      29.717 &    0.325  &   0.016 &    0.005  &          &          \\
 81  &    13:29:54.543 &   47:12:1.225  &    0.081   &             &     2.938   &     0.476 &    2294.173 &   85.862  &   0.417 &           &   0.352  &   6.636  \\
     &                 &                &            &             &     0.019   &     0.006 &      43.555 &    0.559  &   0.019 &           &          &          \\
 82  &    13:29:54.012 &  47:12:14.944  &    0.104   &             &     2.892   &     0.506 &    1539.194 &  161.937  &   0.171 &           &   0.232  &   6.569  \\
     &                 &                &            &             &     0.017   &     0.004 &      22.943 &    0.937  &   0.015 &           &          &          \\
 83  &    13:29:53.060 &  47:12:39.595  &    0.170   &             &     2.952   &     0.662 &    5821.643 &  322.414  &   0.489 &           &   0.896  &          \\    
     &                 &                &            &             &     0.035   &     0.009 &     174.125 &    3.783  &   0.031 &           &          &          \\
 84  &    13:29:52.625 &  47:12:50.828  &    0.203   &     0.195   &     2.974   &     1.831 &    2741.855 &  127.118  &   0.602 &    8.640  &   0.425  &   6.690  \\
     &                 &                &            &     0.022   &     0.125   &     0.179 &     127.577 &    5.345  &   0.060 &    0.015  &          &          \\
 85  &    13:29:59.370 &  47:10:43.181  &    0.283   &             &     2.881   &     0.827 &     199.402 &   40.761  &   0.114 &           &   0.030  &          \\    
     &                 &                &            &             &     0.085   &     0.042 &      17.899 &    1.196  &   0.090 &           &          &          \\
 86  &    13:29:52.053 &  47:11:37.357  &    0.026   &     0.308   &     3.012   &     1.234 &    3654.845 &   16.098  &   0.797 &    8.560  &   0.574  &   9.347  \\
     &                 &                &            &     0.020   &     0.012   &     0.009 &     175.295 &    0.065  &   0.046 &    0.007  &          &          \\
 87  &    13:29:49.087 &  47:11:59.316  &    0.129   &             &     2.994   &     0.828 &    3401.924 &  121.463  &   0.706 &           &   0.531  &          \\    
     &                 &                &            &             &     0.033   &     0.011 &     305.322 &    1.321  &   0.086 &           &          &          \\
 88  &    13:29:58.275 &   47:14:8.694  &    0.467   &             &     2.906   &     0.974 &    1355.002 &  140.732  &   0.245 &           &   0.205  &          \\    
     &                 &                &            &             &     0.036   &     0.013 &      81.642 &    1.760  &   0.058 &           &          &          \\
 89  &    13:29:55.726 &   47:14:1.251  &    0.422   &     0.188   &     2.928   &     1.080 &    2022.547 &  124.729  &   0.361 &    8.596  &   0.309  &   7.569  \\
     &                 &                &            &     0.012   &     0.031   &     0.013 &      58.822 &    1.317  &   0.029 &    0.007  &          &          \\
 90  &    13:29:52.544 &  47:13:51.953  &    0.385   &             &     2.907   &     1.016 &     428.078 &   70.205  &   0.251 &           &   0.065  &   7.277  \\
     &                 &                &            &             &     0.026   &     0.012 &      18.402 &    0.624  &   0.042 &           &          &          \\
 91  &    13:29:47.131 &  47:13:36.161  &    0.387   &     0.142   &     2.928   &     0.980 &     746.196 &   59.876  &   0.364 &    8.612  &   0.114  &   6.934  \\
     &                 &                &            &     0.011   &     0.022   &     0.011 &      22.033 &    0.447  &   0.029 &    0.008  &          &          \\
 92  &    13:29:56.323 &  47:14:10.713  &    0.454   &     0.686   &     2.903   &     1.292 &     371.324 &   91.463  &   0.228 &    8.493  &   0.056  &          \\    
     &                 &                &            &     0.038   &     0.059   &     0.030 &      25.222 &    1.870  &   0.067 &    0.009  &          &          \\
 93  &    13:29:55.437 &   47:14:8.488  &    0.441   &     0.424   &     2.904   &     1.348 &     288.190 &   79.253  &   0.236 &    8.542  &   0.044  &          \\    
     &                 &                &            &     0.030   &     0.040   &     0.020 &      22.310 &    1.082  &   0.075 &    0.010  &          &          \\
 94  &    13:29:50.541 &  47:13:56.032  &    0.405   &     0.298   &     2.914   &     1.130 &    1018.263 &  148.560  &   0.287 &    8.557  &   0.155  &          \\    
     &                 &                &            &     0.016   &     0.038   &     0.015 &      46.249 &    1.938  &   0.045 &    0.007  &          &          \\
 95  &    13:29:47.567 &  47:13:48.465  &    0.414   &     0.360   &     2.911   &     1.164 &     464.696 &   96.794  &   0.274 &    8.543  &   0.071  &          \\    
     &                 &                &            &     0.025   &     0.047   &     0.022 &      32.692 &    1.552  &   0.068 &    0.010  &          &          \\
 96  &    13:29:46.252 &  47:13:45.156  &    0.425   &     0.394   &     2.865   &     1.082 &     754.867 &  237.374  &   0.026 &    8.529  &   0.113  &          \\    
     &                 &                &            &     0.012   &     0.039   &     0.014 &      25.828 &    3.249  &   0.035 &    0.005  &          &          \\
 97$^{a}$  &    13:29:52.698 &  47:11:42.617  &    0.001   &     3.168   &     2.915   &     5.522 &    2728.201 &    8.477  &   0.294 &    8.485  &   0.415  &   9.129  \\
     &                 &                &            &     0.030   &     0.013   &     0.015 &      95.252 &    0.037  &   0.033 &    0.003  &          &          \\
 98  &    13:29:54.712 &  47:11:58.629  &    0.080   &             &     2.925   &     0.570 &    1961.074 &   85.061  &   0.348 &           &   0.299  &   8.409  \\
     &                 &                &            &             &     0.018   &     0.007 &      55.911 &    0.533  &   0.028 &           &          &          \\
 99$^{a}$  &    13:29:52.698 &  47:11:42.617  &    0.001   &     4.702   &     3.063   &     6.917 &   19719.426 &   10.870  &   1.058 &    8.465  &   3.151  &   9.568  \\
     &                 &                &            &     0.061   &     0.016   &     0.020 &    1608.681 &    0.057  &   0.078 &    0.008  &          &          \\
100  &     13:30:4.464 &  47:12:17.279  &    0.395   &             &     2.943   &     0.015 &     258.615 &   21.566  &   0.439 &           &   0.040  &   8.327  \\
     &                 &                &            &             &     0.024   &           &      11.683 &    0.173  &   0.043 &           &          &          \\
101  &     13:30:7.379 &  47:11:29.255  &    0.481   &             &     2.914   &     1.010 &     141.177 &   24.420  &   0.288 &           &   0.021  &   6.681  \\
     &                 &                &            &             &     0.059   &     0.031 &      10.425 &    0.498  &   0.073 &           &          &          \\
102  &     13:30:8.280 &  47:12:49.963  &    0.542   &     0.321   &     2.928   &     1.257 &     214.433 &   26.029  &   0.362 &    8.560  &   0.033  &   6.833  \\
     &                 &                &            &     0.031   &     0.038   &     0.022 &      12.358 &    0.339  &   0.056 &    0.011  &          &          \\
103  &     13:30:1.117 &  47:11:39.335  &    0.274   &     0.068   &     2.983   &     0.776 &    2767.862 &   57.473  &   0.649 &    8.658  &   0.431  &   8.297  \\
     &                 &                &            &     0.005   &     0.015   &     0.005 &      40.883 &    0.283  &   0.015 &    0.007  &          &          \\
104  &    13:29:50.819 &  47:09:18.394  &    0.432   &     0.497   &     2.947   &     1.105 &     156.914 &   13.075  &   0.461 &    8.507  &   0.024  &   8.272  \\
     &                 &                &            &     0.043   &     0.035   &     0.024 &       9.190 &    0.156  &   0.057 &    0.010  &          &          \\
105  &    13:29:53.701 &  47:08:38.541  &    0.549   &     0.285   &     2.873   &     1.114 &     178.119 &   96.303  &   0.069 &    8.561  &   0.027  &   6.734  \\
     &                 &                &            &     0.024   &     0.064   &     0.025 &      12.114 &    2.160  &   0.068 &    0.010  &          &          \\
106  &    13:29:44.304 &  47:09:58.178  &    0.410   &     0.135   &     2.894   &     0.874 &    1015.928 &  109.280  &   0.181 &    8.608  &   0.153  &   7.208  \\
     &                 &                &            &     0.007   &     0.020   &     0.007 &      23.571 &    0.772  &   0.023 &    0.005  &          &          \\
107  &    13:29:49.885 &  47:11:43.551  &    0.092   &     0.150   &     3.012   &     0.638 &    1593.361 &   20.881  &   0.800 &    8.565  &   0.250  &   9.254  \\
     &                 &                &            &     0.017   &     0.016   &     0.008 &      71.903 &    0.114  &   0.043 &    0.011  &          &          \\
108  &    13:29:51.830 &  47:12:58.038  &    0.227   &     0.211   &     2.967   &     0.685 &     363.291 &   18.167  &   0.564 &    8.541  &   0.056  &   8.671  \\
     &                 &                &            &     0.019   &     0.023   &     0.012 &      13.992 &    0.138  &   0.037 &    0.009  &          &          \\
109  &    13:29:38.796 &  47:12:12.349  &    0.464   &             &     2.947   &     1.137 &      86.684 &   14.268  &   0.463 &           &   0.013  &          \\   
     &                 &                &            &             &     0.068   &     0.042 &       9.764 &    0.327  &   0.109 &           &          &          \\
110  &    13:29:46.681 &  47:12:32.509  &    0.249   &     0.118   &     2.948   &     0.830 &     729.011 &   38.824  &   0.465 &    8.614  &   0.112  &   8.443  \\
     &                 &                &            &     0.014   &     0.019   &     0.008 &      22.820 &    0.255  &   0.030 &    0.012  &          &          \\
111  &     13:30:1.260 &  47:13:41.159  &    0.445   &     0.507   &     2.929   &     1.429 &     209.164 &   19.784  &   0.369 &    8.530  &   0.032  &   6.781  \\
     &                 &                &            &     0.039   &     0.032   &     0.021 &      11.255 &    0.213  &   0.052 &    0.009  &          &          \\
112  &     13:30:2.161 &  47:12:56.994  &    0.376   &     0.252   &     2.924   &     1.093 &     502.357 &   30.281  &   0.340 &    8.569  &   0.077  &   8.681  \\
     &                 &                &            &     0.015   &     0.020   &     0.010 &      19.931 &    0.205  &   0.038 &    0.007  &          &          \\
113  &     13:30:6.731 &  47:14:20.010  &    0.648   &     0.326   &     2.934   &     0.846 &     191.527 &   11.331  &   0.394 &    8.522  &   0.029  &   6.995  \\
     &                 &                &            &     0.021   &     0.024   &     0.016 &       5.679 &    0.094  &   0.029 &    0.007  &          &          \\
\enddata
\tablecomments{(1): object number. (2)-(3): equatorial coordinate (J2000). (4): scaled galactocentric distance, where $R$ is the galactocentric distance and $R_{25}$ is the apparent major isophotal radius. (5)-(7): relative fluxes of [{\OIII}]$\lambda$5007,[\NII]$\lambda$6583 and {\Ha} that are normalized to the {\Hb} flux.  (8): {\Hb} flux in unit of $10^{-17}\mathrm{erg~s^{-1}~cm^{-2}}$. (9): equivalent width of {\Ha} line in {\AA} . (10): gas-phase reddening in mag. (11): metallicity in dex. (12): SFR density in $\rm M_{\sun} yr^{-1} kpc^{-2}$. (13): logarithmic stellar age in year. Each row is followed by another row presenting the error.}
\tablenotetext{a}{AGN region.}
\end{deluxetable*}
\end{longrotatetable}



\end{document}